\documentclass[prd,showpacs,nofootinbib,superscriptaddress,preprintnumbers,preprint,tightenlines]{revtex4}

\usepackage[mathscr]{eucal}
\usepackage[latin1]{inputenc}
\usepackage{amsmath}
\usepackage{amsfonts}
\usepackage{amssymb}
\usepackage{pictex}
\usepackage{natbib,hyperref}
\usepackage{graphicx}

\def\be{\nopagebreak[3]\begin{equation}}
\def\ee{\end{equation}}
\def\ba{\nopagebreak[3]\begin{eqnarray}}
\def\ea{\end{eqnarray}}

\newcommand{\teta}{\rlap{\lower2ex\hbox{$\,\tilde{}$}}\eta{}}

\def\lp{{\ell}_{\rm Pl}}
\def\q{\mathring{q}}

\newcommand{\rcr}{\rho_{\mathrm{crit}}}

\newcommand{\heff}{{\cal H}_{\mathrm{eff}}}
\newcommand{\calh}{{\cal H}}

\newcommand{\p}{\partial}

\newcommand{\f}{\frac}

\usepackage{enumerate}

\usepackage{colordvi}
\newcounter{mnotecount}[section]

\newcommand{\comment}[1]{}

\def\f{\frac}

\def\epsilon{\varepsilon}

\begin{document}
\preprint{\vbox{\baselineskip=12pt \rightline{PI-QG-141}
}}

\title{Non-singular Ekpyrotic/Cyclic model in Loop Quantum Cosmology}

\author{Thomas Cailleteau}
\affiliation{Laboratoire de Physique Subatomique et de Cosmologie, UJF,CNRS-IN2P3-INPG
53 Avenue des Martyrs, 38026 Grenoble Cedex, France
}
\affiliation{Institute of Cosmology and Gravitation, University of Portsmouth, 
Portsmouth P01 2EG, United Kingdom}
\author{Parampreet Singh}
\email{psingh@perimeterinstitute.ca}
\affiliation{Perimeter Institute for
Theoretical Physics, 31 Caroline Street North, Waterloo, Ontario
N2L 2Y5, Canada}
\author{Kevin Vandersloot}
\affiliation{Institute of Cosmology and Gravitation, University of Portsmouth, 
Portsmouth P01 2EG, United Kingdom}

\begin{abstract}

We study the role of non-perturbative quantum gravity effects in the Ekpyrotic/Cyclic model using the effective framework of loop quantum cosmology in the presence of anisotropies. We show that quantum geometric modifications to the 
dynamical equations near the Planck scale as understood in the quantization of Bianchi-I spacetime in loop quantum cosmology lead to the resolution of classical singularity and result in a non-singular transition of the universe from the contracting to the expanding branch. In the Planck regime, the universe undergoes multiple small bounces and the anisotropic shear remains bounded throughout the evolution. A novel feature, which is absent for  isotropic models, is a natural turn around of the moduli field from the negative region of the potential leading to a cyclic phenomena as envisioned in the original paradigm. Our work suggests that incorporation of  quantum gravitational effects in the Ekpyrotic/Cyclic model may  lead to a viable scenario without any violation of the null energy condition.

\end{abstract}

\maketitle

\section{Introduction}

One of the most intriguing issues in cosmology concerns with the state of the universe in the earliest epoch of its evolution. It is this phase where answers to some of the most difficult  questions in conventional cosmology are hidden which includes understanding 
the initial conditions of our universe. It is difficult to understand the latter via 
cosmological models based on classical general relativity (GR) since it 
 break downs at very large spacetime curvatures and predicts an initial 
singularity. Therefore, any viable description of the history of our universe must 
necessarily have inputs from a framework beyond GR. Not surprisingly, one expects that any such model must capture quantum aspects of gravity in order to be successful.
Though a treatment based on a full theory of quantum gravity is beyond the scope at the present stage, nevertheless useful insights have been obtained on the nature of the universe at Planck scale in various frameworks. In one of the earliest works in this direction, Wheeler 
showed that the spacetime near the 
initial singularity resembles a quantum foam \cite{geons} (a picture shared with the analysis of 
quantization of conformal modes in gravity \cite{pn:nature}).  In recent years 
a different picture seems to emerge in some models based on non-perturbative canonical quantum gravity. In loop quantum cosmology (see Ref.\cite{ashtekar:lqc_review} for an introductory review) a quantization of symmetry reduced spacetimes based on background independent loop quantum gravity, backward evolution of states for different stress-energy contents, show existence of a bounce near the Planck scale to a classical contracting  branch \cite{aps:prl,aps:mu0,aps:improved}. The quantum bounce is non-singular and allows a unitary evolution across the ``big bang'' {\it without any violation of null energy condition (NEC)}.
These results indicate that the universe may have a 
``pre big bang'' branch which plays an important role in the history and the fate of the universe.

Interestingly, though loop quantum cosmology predicts existence of a contracting universe preceding ours, the idea of a pre big bang branch of the universe is not new. Various paradigms in theoretical cosmology have conjectured an existence of such a pre big bang branch and existence of a non-singular bounce is the most crucial element for them to be viable.
As an example, pre big bang branch is envisioned in Ekpyrotic/Cyclic model paradigm \cite{kst:designingcyclic,st:cyclic1,st:cyclic2} which is  also 
considered as an alternative to the inflationary scenarios (see Ref. \cite{lehners_review} for a review). It is based on the  inputs from M-theory where it is envisioned that the universe undergoes cycles of expansion and contraction governed 
by the inter-brane dynamics of two boundary branes in a five dimensional bulk.\footnote{Based on string theory, idea of a bouncing universe has also been extensively studied  in Pre-Big Bang models \cite{pbb_review} and more recently in string gas cosmologies \cite{sgas_bounce}.} Our 
observable universe is hypothesized to be confined on one of the boundary branes which 
interacts only gravitationally with the other brane. Attraction between the branes leads to their collision, an event which corresponds to a big bang or big crunch singularity for the observable universe.

The transition from the contracting to the expanding branch when the branes collide and separate is a tricky
 issue  in the Ekpyrotic/Cyclic models. Though insights on this problem from the 
 bulk   spacetime perspective 
have been gained  (see Refs. \cite{cyclic_pert1,tch:adscft1,cht:adscft2}), its solution remains elusive in 
the 4-dimensional cosmological picture. In its absence  a viable non-singular cosmological model of the early universe is 
difficult to construct. Insights on obtaining a non-singular transition in the Ekpyrotic/Cyclic model is expected from understanding the role of non-perturbative quantum gravitational effects which are well understood in loop quantum gravity. In particular a lot of progress has been made in recent years on understanding resolution of singularities in the framework of loop quantum cosmology \cite{ashtekar:lqc_overview,ashtekar:lqc_review,bojowald:livingrev,singh:review1}.
This leads to a natural question (which is the focus of this work): Is it possible to obtain a non-singular transition in the Ekpyrotic/Cyclic model using inputs from loop quantum cosmology? 

Loop quantum cosmology is a 
canonical non-perturbative quantization of homogeneous spacetimes based on Dirac's method of constraint quantization. The gravitational part of the classical phase space is labeled by the Ashtekar variables: SU(2) connection $A^i_a$ (which in a cosmological setting is proportional to rate of change of scale factor at the classical level) and conjugate triad $E^a_i$ (proportional to the square of the scale factor). The elementary variables used for quantization in loop quantum cosmology are the holonomies of the connection components and triads. Quantization proceeds with expressing the classical Hamiltonian constraint, consisting of the field strength tensor of the Ashtekar connection, in terms of 
elementary variables. 
The resulting quantum constraint captures the underlying discreteness in quantum geometry (one of the predictions of loop quantum gravity) and turns out to be uniformly discrete in volume. This has been shown to be a property of   isotropic models (for all values of curvature index $k$) \cite{aps:improved,apsv:closed,skl:closed,vandersloot:open,szulc:open,acs:slqc} and Bianchi-I spacetimes \cite{aw:bianchi1} where one has rigorous analytical control and 
extensive numerical simulations  have been performed.

Some  of the novel features of loop quantization can be described as follows. Let us consider a isotropic universe filled with a massless scalar field. Take a state which is peaked at the classical trajectory at late times and evolve it with the loop quantum constraint towards the big bang. It turns out that the state remains peaked at the classical trajectory till spacetime curvature $(R)$ is approximately $1\%$ of the Planck value, however on further evolution there are significant departures from the classical theory. Instead of following the classical trajectory till the big bang as in the Wheeler-DeWitt theory, evolution leads to a bounce of the universe 
when energy density of the universe reaches a critical value, $\rcr = 0.41 \rho_{\mathrm{Pl}}$ \cite{aps:improved}. After the bounce the state then evolves further to peak on the classical trajectory for a contracting universe. These turn out to be 
robust features of the theory. Using an exactly solvable model it can be proved that the bounce is a property {\it for all the states} in the physical Hilbert space and there exists a supremum for the energy density operator given by the above value \cite{acs:slqc}. Moreover, the fluctuations are bounded across the bounce and the state which is semi-classical at late times post bounce is also semi-classical at early times pre bounce \cite{cs:recall}. It is important to note that existence of bounce does not require any fine tuning of initial conditions or a special choice of matter. 
In fact for the above case, matter satisfies stiff equation of state throughout the evolution. 
Existence of bounce and non-singular evolution has also been shown for scalar field with cosmological constant (for  positive \cite{ap:positivecc} as well as negative values \cite{bp:negativecc}) and the inflationary potential \cite{aps:inflaton}. Another important feature of loop quantum cosmology is that it turns out that there exists a unique consistent loop quantization for the isotropic spacetimes as well as the Bianchi-I spacetime \cite{cs:unique,cs:geom}.

Interestingly, it is possible to write an effective Hamiltonian in loop quantum cosmology with a resulting dynamics which approximates the underlying quantum dynamics to an excellent accuracy. The effective Hamiltonian can be derived using coherent state techniques for different matter sources \cite{taveras:fried,willis:thesis,singh:effective}. Assuming that effective dynamics is valid for arbitrary matter (an assumption which turns out to be true for various cases), it is possible to prove that isotropic flat loop quantum cosmology is generically non-singular and geodesically complete \cite{singh:nonsingular}. It resolves all known types of cosmological singularities. Similar results are expected for curved and anisotropic models. Results obtained in the full quantum theory, numerical simulations and the effective theory strongly indicate that resolution of cosmological singularities is natural in loop quantum cosmology.

The availability of an effective description leads us to explore a viable non-singular bouncing model in the effective description of loop quantum cosmology with the Ekpyrotic/Cyclic model potential. This  serves as a first step to investigate the role of non-perturbative quantum gravity effects as derived in loop quantum gravity to the Ekpyrotic/Cyclic models. For the case when anisotropies are absent, such an analysis was performed earlier \cite{svv:cyclic}. It was shown that a non-singular bounce is generically possible for the 
flat isotropic Ekpyrotic/Cyclic model in loop quantum cosmology. Despite this a viable model was not possible due to lack of a turn around of the 
moduli field in the epoch when the branes collide and separate, inhibiting a cyclic phenomena. It   was shown that unless the 
Ekpyrotic/Cyclic model potential becomes positive when the singularity is approached  it is not possible 
for the moduli field to turn around in the process of transition from contracting to the  expanding branch (or vice versa) and lead to cycles.\footnote{Such a modification to the Cyclic model potential has also been suggested in Refs. \cite{ffkl:negative,Lehners}.}

However in  a realistic universe anisotropies are always present, even if their strength be very small. Hence it 
is important to analyze whether the limitations pointed out in the analysis of Ref.\cite{svv:cyclic} were artifacts of the 
assumption of pure isotropy. It is pertinent to ask whether 
there exists a viable non-singular Ekpyrotic/Cyclic model for effective loop quantum dynamics incorporating anisotropic properties of spacetime? As we will show the answer turns out to be positive.  We will see that the mere presence of shear term in the cosmological dynamics leads to a turn around of the moduli field from the negative region of the potential. Instead of a single bounce of the scale factor, the anisotropic effective dynamics obtained from the loop quantization of Bianchi-I model exhibits bounces for each of the directional scale factors. Numerical simulations show that the turn around of the moduli occurs in the middle of the transition of the mean scale factor from the contracting to the expanding branch. Further, the anisotropic shear remains bounded throughout the evolution. Thus a non-singular transition from the contracting to the expanding phase with a turn around of the moduli as envisioned in the 
Ekpyrotic/Cyclic model paradigm is achieved in the effective loop quantum dynamics. 
This is the main result of our analysis.

This paper is organized as follows. In the next section we re-visit the classical theory of Bianchi-I model in the Ashtekar variables. We describe the way classical generalized Friedman equation for an anisotropic model can be derived in a Hamiltonian treatment. In Sec. III we consider the effective Hamiltonian of the Bianchi-I model for the loop quantization performed in Ref. \cite{aw:bianchi1}. This quantization for Bianchi-I model has recently been shown to be the only consistent choice which leads to a bounded shear and expansion factors \cite{cs:geom}. Using Hamilton's equations we derive the dynamical equations for connections and triad components (the dynamical equations for matter such as the Klein-Gordon equation are not changed by quantum gravitational effects) and highlight some of the properties of the effective theory. In Sec IV, we use the effective dynamics to analyze the potential in the Ekpyrotic/Cyclic model. Using numerical techniques we study the evolution of the moduli field focusing in particular on the transition from the contracting branch to the expanding branch. We show that a non-singular turn around of the scale factor along with the same for the moduli field is possible in the effective 4-dimensional description without any extra inputs. 
We summarize the results and discuss open issues in the concluding section.

\section{Bianchi-I model: Classical theory in Ashtekar variables}

The Bianchi-I spacetime is one of the simplest examples of spacetimes with anisotropies. It has vanishing intrinsic curvature and unlike Bianchi-IX model, the classical dynamics does not exhibit BKL behavior as the singularity is approached. 
The isotropic limit of this spacetime is $k=0$ FRW cosmological spacetime.
We consider a homogeneous  Bianchi-I anisotropic spacetime with a manifold $\Sigma \times \mathbb{R}$ where $\Sigma$ is topologically flat $(\mathbb{R}^3)$. The spatial manifold is non-compact and in order to define the symplectic structure and formulate a Hamiltonian theory, it is necessary to introduce a fiducial cell ${\cal V}$. The cell ${\cal V}$ has fiducial volume $V_o = l_1 l_2 l_3$ with respect to the fiducial metric $\q_{ab}$ endowed on the spatial manifold. Here $l_i$ refer to the coordinate lengths of the each side of the 
fiducial cell.\footnote{As should be expected from any consistent treatment, physical predictions of this model are insensitive to the choice of $l_i$ and one could choose these to be unity. We do not restrict to this choice in order to stress the independence of 
physics in the classical and especially the effective theory on the fiducial cell. The latter feature is not shared by alternative quantizations of Bianchi-I model in loop quantum cosmology.}

 Due to the homogeneity of the Bianchi-I spacetime,  the Ashtekar variables take a simple form. The matrix valued connection $A^i_a$ and triad $E^a_i$ can be expressed as $c_i$ and $p_i$ respectively (where $i = 1,2,3$) \cite{cv:bianchi1}. The 
canonical conjugate phase space variables satisfy
\be
\{c_i, p_j\} = 8 \pi G \gamma \delta_{ij} ~
\ee
where $\gamma = 0.2375$ is the Barbero-Immirzi parameter. 
The triad $p_I$ are related to the three scale factors $a_I$ of the Bianchi-I metric
\be
d s^2 = -N^2 \, d t^2 + a_1^2 \, d x^2 + a_2^2 d y^2 + a_3^2 d z^2
\ee
as
\be\label{triadsf}
|p_1| = l_2 l_3 \, a_2 a_3, ~~~ |p_2| = l_1 l_3 \, a_1 a_3, ~~~ |p_3| = l_2 l_3 \,a_2 a_3 ~
\ee
where the modulus sign arises because of orientation of the triads (and is suppressed in the following).

The only non-trivial constraint to be solved in this model is the Hamiltonian constraint which when expressed 
in terms of $c_i$ and $p_i$ takes the form:
\be\label{clH}
{\cal H}_{\mathrm{cl}} = -\f{N}{8 \pi G \gamma^2 V}{(c_1 p_1 \, c_2 p_2 + c_3 p_3 \, c_1 p_1 + c_2 p_2 \, c_3 p_3)} + {\cal H}_{\mathrm{matt}} ~
\ee
where ${\cal H}_{\mathrm{matt}}$ is the matter Hamiltonian which may describe perfect fluid and/or scalar fields.  
Physical solutions are obtained by the vanishing of the Hamiltonian constraint: ${\calh}_{\mathrm{cl}} \approx 0$. Equations of motions for the phase space variables are 
obtained by solving Hamilton's equations:
\be\label{pdot}
\dot p_i = \{p_i, {\cal H}_{\mathrm{cl}}\} = - 8 \pi G \gamma \, \f{\p \calh}{\p c_i}
\ee
and 
\be\label{cdot}
\dot c_i = \{c_i, {\cal H}_{\mathrm{cl}}\} =  8 \pi G \gamma \, \f{\p \calh}{\p p_i} ~.
\ee
If we choose the lapse function $N=1$, then using (\ref{clH}) and (\ref{cdot}) we obtain
\be\label{ceq}
c_i = \gamma l_i \dot a_i = \gamma l_i \, H_i a_i~
\ee
where $H_i \equiv \dot a_i/a_i$ is the Hubble rate in the i-th direction.

Similarly we can obtain the dynamical equations for the matter degrees of freedom. As an example, in case ${\cal H}_{\mathrm{matt}}$ corresponds to a minimally coupled scalar field $\phi$ with momentum $p_\phi$ (satisfying $\{\phi, p_\phi\} = 1$) and potential $V(\phi)$, the dynamical equations take the standard form
\be\label{mateq1}
\dot \phi = \f{\p}{\p p_\phi}\,  {\cal H}_{\mathrm{matt}} ~~~ \mathrm{and} ~~~ \dot  p_\phi = - \f{\p}{\p \phi}\,  {\cal H}_{\mathrm{matt}} ~.
\ee
Taking the second derivative of $\phi$ and using $\dot p_\phi$, we obtain the  standard Klein-Gordon equation for $\phi$:
\be\label{mateq2}
\ddot \phi \, + \, \sum_i H_i \,  \phi = - \p_\phi V(\phi) ~.
\ee

Before we study the properties of dynamical equations it is useful to note the behavior of the classical Hamiltonian constraint under one of the underlying freedoms in the framework. This is to do with the change of the shape of the fiducial cell. Let us consider this change as:  $(l_1, l_2, l_3) \rightarrow (l_1', l_2', l_3')$. Under this change $V \rightarrow l_1' l_2' l_3' V$ and  $c_1, p_1$ (and similarly other components) transform as 
\be\label{cptrans}
c_1 \rightarrow c_1' = l_1' c_1 ~, ~~~~ p_1 \rightarrow p_1' = l_2' l_3' p_1 ~.
\ee
Thus, the gravitational part of the constraint transforms as ${\cal H}_{\mathrm{grav}} \rightarrow l_1' l_2' l_3' {\cal H}_{\mathrm{grav}}$. It can be shown that the matter part of the constraint also transforms in the same way. Thus, ${\cal H}_{\mathrm{class}}/V$ is invariant under arbitrary change in shape of the fiducial cell.

Solving the classical constraint ${\cal H}_{\mathrm{cl}} \approx 0$, dividing by the 
 total volume
$ V = V_o (a_1 a_2 a_3) = (p_1 p_2 p_3)^{1/2} $ and using (\ref{ceq})
 we obtain the following equation relating directional Hubble rates with energy density
\be \label{bf1}
H_1 H_2 + H_2 H_3 + H_3 H_1 = 8 \pi G \, \f{\cal{H}_{\mathrm{matt}}}{V} = 8 \pi G \, \rho ~
\ee
where $\rho$ is the energy density of the matter component: $\rho = {\cal H}_{\mathrm{matt}}/V$.  As expected, using (\ref{cptrans}) one finds that Hubble rates and energy density are invariant under $(l_1, l_2, l_3) \rightarrow (l_1', l_2', l_3')$.\\

An interesting property of equations of motion for $p_i$ and $c_i$ arises for matter 
with vanishing anisotropic stress. It can then be shown that 
$(p_i c_i - p_j c_j)$ is a constant of motion satisfying \cite{cv:bianchi1}, i.e.
\be\label{consteq}
p_i c_i - p_j c_j = V (H_i - H_j) = \gamma V_o \alpha_{ij}
\ee
where $\alpha_{ij}$ is a constant antisymmetric matrix. 

The directional Hubble rates can be considered to be the diagonal elements of an 
expansion matrix $H_{i j}$. The trace of this matrix is related to the
expansion rate of geodesics in this spacetime as
\be
\theta = \f{1}{3} (H_1 + H_2 + H_3) = \f{1}{a} \f{d a}{d t}
\ee
where $a$ identified as the mean scale factor for our anisotropic model:
\be
a = (a_1 a_2 a_3)^{1/3} ~.
\ee

The trace-free part of the expansion matrix leads to the shear term $\sigma_{i j}$:
\be
\sigma_{i j} = H_{i j} - \theta \delta_{ij}
\ee
and defines the  shear scalar $\sigma^2 \equiv \sigma_{\mu \nu} \sigma^{\mu \nu}$ given by 
\begin{eqnarray}\label{sigma1}
\sigma^2 &=& \nonumber  \sum_{i=1}^3 (H_i - \theta)^2 ~~ = ~~\f{1}{3} \left((H_1 - H_2)^2 + (H_2 - H_3)^2 + (H_3 - H_1)^2\right)
\\&=& \f{1}{3 a^6} \left(\alpha_{12}^2 + \alpha_{23}^2 + \alpha_{31}^2\right)
\end{eqnarray}
where to obtain the last expression we have used Eq.(\ref{consteq}).

The  generalized Friedman equation for the mean scale factor can be obtained by 
considering the mean Hubble rate 
\be
H = \f{\dot a}{a} = \f{1}{3} \sum_{i=1}^3 H_i ~,
\ee
which yields
\be
H^2 =  \f{1}{3} (H_1 H_2 + H_2 H_3 + H_3 H_1) ~~ + ~~ \f{1}{18} \left((H_1 - H_2)^2 + (H_2 - H_3)^2 + (H_3 - H_1)^2\right)\nonumber .\\
\ee
On using Eqs.(\ref{bf1}) and (\ref{sigma1}) we obtain 
\be\label{clf}
H^2 =  \f{8 \pi G}{3} \rho \, + \, \f{\Sigma^2}{a^6} ~.
\ee
with
\be\label{sigmascalar}
\Sigma^2 \equiv  \f{1}{6} \, \sigma^2 a^6 ~.
\ee
From (\ref{sigma1}) it follows that the shear scalar $\Sigma$ is  a constant of motion in the classical theory. As we will show later, this feature does not hold in the loop quantization where only at the classical scales $\Sigma$ approaches a constant value.

Analysis of the generalized Friedman equation (\ref{clf}) immediately shows that the singularity is inevitable in the classical theory as the scale factor approaches zero. 
It is important to note that in the classical theory anisotropic term dominates on approach to singularity unless $\rho$ corresponds to matter with equation of state $w > 1$ i.e. with an equation of state of an ultra-stiff fluid. In the case when matter has equation of state $w < 1$, the energy density grows slower than anisotropic term as $a(t) \rightarrow 0$. We will discuss later that in the Ekpyrotic/Cyclic model the equation of state $w \gg 1$ during the ekpyrosis phase (when the moduli is in the steep negative region of the potential). This causes anisotropic term to be subdued in subsequent evolution. Thus, as the universe approaches classical big bang/crunch singularities in this model one expects anisotropies to become very small \cite{glps:smooth}.

\section{Effective Dynamics}

In the loop quantization the elementary variables are the holonomies of the connection and the flux of the triad (related by a constant for the present case). Elements of the holonomies are of the form 
$\exp(i \mu c_i)$ where $\mu$ labels the edge length along which a holonomy is computed.
The classical Hamiltonian constraint is expressed in terms of holonomies and triads and then quantized. The resulting quantum constraint is
generically non-singular and the picture at the Planck scale in Bianchi-I model turns out to be similar to the one in the isotropic model\cite{aw:bianchi1}. One can also derive the effective Hamiltonian using similar techniques as used for the 
isotropic model \cite{taveras:fried} and for $N=1$ is given by\footnote{In order to compare with earlier works (for eg. equations in Ref.\cite{cv:bianchi1,cs:geom}), in our convention $\bar \mu_i$ correspond to often used $\bar \mu_i'$.}
\be\label{effham}
\heff =  - ~\f{1}{8 \pi G \gamma^2 V}\left(\f{\sin(\bar \mu_1 c_1)}{\bar \mu_1} \f{\sin(\bar \mu_2 c_2)}{\bar \mu_2}  p_1 p_2 + \mathrm{cyclic} ~~ \mathrm{terms}\right) ~~ + ~~ {\cal H}_{\mathrm{matt}}~  \\
\ee
where
\be \label{mub1}
\bar \mu_1 = \lambda \sqrt{\f{ p_1 }{p_2 p_3}}, ~~~ \bar \mu_2 = \lambda \sqrt{\f{p_2}{p_1 p_3}}, ~~~ \mathrm{and} ~\bar \mu_3 = \lambda \sqrt{\f{p_3}{p_1 p_2}} ~ .
\ee
Here $\Delta$ arises due to regularization of the field strength of the connection by the underlying quantum geometry. Its value is given by \cite{aw:entropy}
\be
\lambda^2 = 4 \sqrt{3} \pi \gamma \lp^2 ~.
\ee
It is to be noted that the effective dynamics resulting from the above Hamiltonian is invariant under the choice of the fiducial cell. 
In particular this is the only known loop quantization of Bianchi-I spacetime which is independent of the shape of the fiducial cell \cite{cs:geom}. Comparing (\ref{effham}) and (\ref{clH}) we notice that the change from classical to effective theory  only comes in the form of replacement of connection components $c_I$ with $\sin(\bar \mu_I c_I)/\bar \mu_I$. Under the transformation: $(l_1, l_2, l_3) \rightarrow (l_1', l_2', l_3')$, $\bar \mu_I \rightarrow \bar \mu_I' = (1/l_I') \bar \mu_I$. Hence $\sin(\bar \mu_I c_I)/\bar \mu_I$ transforms in the same way as $c_I$. Thus transformation 
properties remain same as in the classical theory and the 
resultant dynamics and physical predictions are unaffected by the freedom of the choice of the fiducial cell.\footnote{This is in contrast to other proposals for quantization of Bianchi-I spacetimes such as in Ref.\cite{chiou:bianchi1} and those motivated by lattice refinement considerations \cite{lattice:bianchi1}. As emphasized in Ref. \cite{cs:geom}, 
since the effective dynamics in these models is not invariant under the change in shape of the fiducial cell, they lack predictive power. It turns out that in the effective dynamics of these proposals the expansion factor and shear are also unbounded, even if one fixes the fiducial cell.}

Substituting (\ref{mub1}) in (\ref{effham}) and solving for the Hamiltonian constraint $\heff \approx 0$ we obtain
\be
\rho = \f{1}{8 \pi G \gamma^2 \lambda^2} \left(\sin(\bar \mu_1 c_1) \, \sin(\bar \mu_2 c_2)  + \mathrm{cyclic} ~~ \mathrm{terms}\right) ~.
\ee
Since $\sin(\bar \mu_i c_i)$ are bounded functions, above equation implies that the 
 energy density can never diverge in the 
effective loop quantum cosmology. The maximum value of the terms in the parenthesis determines the upper bound for the energy density:
\be\label{rhomax}
\rho \leq \rho_{\mathrm{max}} , ~~\rho_{\max} = \f{3}{8 \pi G \gamma^2 \lambda^2} ~.
\ee
The upper bound for the energy density in the Bianchi-I anisotropic model coincides with the value in the isotropic model \cite{aps:improved}.

It should be noted that the matter Hamiltonian ${\cal H_{\mathrm{matt}}}$ in (\ref{effham}) is unmodified from its classical expression, due to which the dynamical equations for matter remain the same as Eqs.(\ref{mateq1}) and (\ref{mateq2}).
Modified dynamical equations can be obtained from the Hamilton's equations 
using (\ref{effham}). As an example, for the triad component $p_1$ we get
\ba\label{dotp1}
\dot p_1 &=& \nonumber  - 8 \pi G \, \f{\partial {\cal H}_{\mathrm{eff}}}{\p c_1} \\
&=& \f{p_1}{\gamma \lambda} \, \cos(\bar\mu_1 c_1) \, \left(\sin(\bar \mu_2 c_2) + \sin(\bar \mu_2 c_2) \right) ~
\ea
(and similarly for $p_2$ and $p_3$).

In order to find the equation for directional Hubble rates we first note that
from (\ref{triadsf}) we get
\be
a_1 = \f{1}{l_1} \, \left(\f{p_2 p_3}{p_1}\right)^{1/2}~, ~~~a_2 = \f{1}{l_2} \, \left(\f{p_3 p_1}{p_2}\right)^{1/2}~ \mathrm{and} ~ ~a_3 = \f{1}{l_3} \, \left(\f{p_1 p_2}{p_3}\right)^{1/2}~.
\ee
Taking their derivatives and using Eq.(\ref{dotp1}) and corresponding equations for
$p_2$ and $p_3$ we obtain
\be\label{eqh1}
H_1 = \f{\dot a_1}{a_1} = \f{1}{2 \gamma \lambda} \left(\sin(\bar \mu_1 c_1 - \bar \mu_2 c_2) + \sin(\bar \mu_1 c_1 - \bar \mu_3 c_3) + \sin(\bar \mu_2 c_2 + \bar \mu_3 c_3) \right) ~.
\ee
Similar equations can be derived for the directional Hubble rates $H_2$ and $H_3$. It is clear from the above equation, unlike in the classical theory the directional Hubble rates in effective loop quantum dynamics are always bounded.

The Hamilton's equation for connection components can be derived in a similar way. As an example, for $c_1$ one obtains 
\ba\label{dotc1}
\dot c_1 &=& \nonumber  8 \pi G \gamma \f{\partial H_{\mathrm{eff}}}{\partial c_1} ~\nonumber \\&=& \f{1}{2 p_1 \gamma \lambda}\Bigg[c_2 p_2 \cos(\bar \mu_2 c_2) (\sin(\bar \mu_1 c_1) + \sin(\bar \mu_3 c_3)) + ~c_3 p_3 \cos(\bar \mu_3 c_3) (\sin(\bar \mu_1 c_1) + \sin(\bar \mu_2 c_2)) \nonumber \\
&&
 \hskip1.5cm - ~ c_1 p_1 \cos(\bar \mu_1 c_1) (\sin(\bar \mu_2 c_2) + \sin(\bar \mu_3 c_3)) - ~\bar \mu_1 p_2 p_3 \bigg[\sin(\bar \mu_2 c_2)\sin(\bar \mu_3 c_3) \nonumber  \\ && \hskip1.5cm + \sin(\bar \mu_1 c_1)\sin(\bar \mu_2 c_2) + \sin(\bar \mu_3 c_3)\sin(\bar \mu_1 c_1) \bigg]
\Bigg] ~  + ~ 8 \pi G \gamma \, \sqrt{\f{p_2 p_3}{p_1}} \, \left(\f{\rho}{2} + p_1 \f{\partial \rho}{\partial p_1}\right) ~. \nonumber \\
\ea

Using $(\ref{dotp1})$ with $(\ref{dotc1})$ and similar equations for othe connection and triad components, it can be shown that matter with vanishing stress satisfies
\be
\f{d}{d t} (p_i c_i - p_j c_j) = 0 ~.
\ee
Thus as in the classical theory $(p_i c_i - p_j c_j)$ is a constant of motion in the effective loop quantum dynamics. However, due to Eq.(\ref{eqh1}), unlike in the classical theory the relation (\ref{ceq}) is no longer 
satisfied. A consequence is that in the effective dynamics 
\be
p_i c_i - p_j c_j \neq V (H_i - H_j) ~.
\ee
In the classical approximation $\bar \mu_i c_i \ll 1$, Eq.(\ref{eqh1}) leads to $c_i \approx \gamma l_i \dot a_i$ which implies $p_i c_i - p_j c_j \approx V (H_i - H_j)$ as we approach the classical scales.\\

We can now evaluate the shear scalar $\sigma^2$ in the effective description of LQC. Since $(H_i - H_j)$ are no longer constant except at the classical scales, we note from Eq.(\ref{sigma1})
that $\sigma^2$ is not a constant in the effective LQC. Using the Hamilton's equations for $p_i$, a straightforward calculation leads to
\ba\label{sigmasqlqc}
\sigma^2 &=& \nonumber \f{1}{3 \gamma^2 \lambda^2} \Bigg[\left(\cos(\bar \mu_3 c_3) (\sin(\bar \mu_1 c_1) + \sin(\bar \mu_2 c_2))  - \cos(\bar \mu_2 c_2) (\sin(\bar \mu_1 c_1) + \sin(\bar \mu_3 c_3))\right)^2 \\ && \nonumber + 
\left(\cos(\bar \mu_3 c_3) (\sin(\bar \mu_1 c_1) + \sin(\bar \mu_2 c_2)) - \cos(\bar \mu_1 c_1) (\sin(\bar \mu_2 c_2) + \sin(\bar \mu_3 c_3))\right)^2 \\ &&  + \left(\cos(\bar \mu_2 c_2) (\sin(\bar \mu_1 c_1) + \sin(\bar \mu_3 c_3))  - \cos(\bar \mu_1 c_1) (\sin(\bar \mu_2 c_2) + \sin(\bar \mu_3 c_3))\right)^2\Bigg] ~.
\ea

This implies that $\Sigma^2$ defined via Eq.(\ref{sigmascalar}) is not a constant of 
motion in the effective loop quantum dynamics. Nevertheless, it is clear that   
$\sigma^2$ is bounded above by a fundamental value in loop quantum cosmology. A detailed  analysis of the behavior of anisotropy and energy density will be done else where. It is worth pointing out some of the notable features of the effective dynamics.  These include:\\ 
(i) Unlike the bounce in the isotropic model, due to presence of anisotropies the energy density at the bounce may be less than $\rho_{\mathrm{max}}$. In fact it is possible for the bounce to occur with $\rho = 0$. In this case bounce occurs purely because of interplay of quantum geometric effects with anisotropy.\\
(ii) The maximum value of anisotropic shear at the bounce is 
\be
\sigma^2_{\mathrm{max}} = \f{4}{3 \gamma^2 \lambda^2} ~.
\ee
(iii) It is also possible for the effective dynamics to saturate Eq.(\ref{rhomax}) at the bounce. However, in that case anisotropy vanishes at the bounce. 

These features exhibit richness of the effective dynamics in the Bianchi-I model. They result due to incorporation of quantum geometry effects in the effective Bianchi-I spacetime. When the components of the spacetime curvature become large, there are significant departures from the classical theory. When these components are small compared to the Planck scale then dynamical equations approximate their classical counterparts and one recovers the classical description. It is important to note that 
the upper bounds on energy density, shear and Hubble rate are direct consequences of 
the underlying quantum geometry. In the classical limit $\lambda \rightarrow 0$, $\rho_{\mathrm{max}}, H_{\mathrm{max}}$ and $\sigma_{\mathrm{max}}$ diverge and the evolution is singular.
The boundedness of these quantities  plays the crucial role to construct non-singular 
Ekpyrotic/Cyclic model as we we discuss in the following section.

\section{Evolution with the Cyclic Potential}

In the previous section we analyzed the effective dynamics of Bianchi-I model in loop quantum cosmology for matter with vanishing anisotropic stress. An important feature of quantum geometry modified  dynamics turns out to be  
 singularity resolution. We showed that the energy density, Hubble rates and shear are always bounded in loop quantum cosmology independent of the equation of state of the matter content. This result is very encouraging for the 
primary question posed in this work: Is there a viable non-singular evolution with the Ekpryotic/Cyclic potential in the 
effective loop quantum cosmology? We now explore the answer to this question.

Let us first briefly recall some of the salient features of Ekpyrotic/Cyclic model which is motivated by M-theory. It is hypothesized that the observable universe is constrained on a visible brane which interacts gravitationally with a shadow brane. The collision between these branes constitutes the big bang/big crunch singularities for the observable universe in the 4-dimensional effective description. The inter-brane separation is determined 
by a moduli field $(\phi)$ and the interaction potential is given by \cite{st:cyclic1}
\be\label{cyclicpot}
V = V_o (1 - e^{-\sigma_1 \phi}) \exp(-e^{-\sigma_2 \phi}) ~
\ee
where $V_o$, $\sigma_1$ and $\sigma_2$ are the parameters of the potential. 

An attractive feature of the Ekpryotic/Cyclic models is the different roles which the 
moduli plays in various epochs. When the branes approach each other, the moduli field 
moves slowly in an  almost flat and positive part of the potential which in the effective 4-dimensional 
description leads  to a period of dark energy domination.  Eventually when the potential energy of the moduli balances the Hubble expansion rate, the moduli field turns around 
and the inter-brane separation starts decreasing leading to a contracting phase in the universe. When the branes approach each other, the moduli potential is steep and negative leading to an ultra-stiff equation of state, i.e. $w > 1$. It turns out that this feature leads to a decrease in the anisotropy of the universe at small scale factors \cite{glps:smooth}. The ultra-stiff equation of state 
is also responsible for producing a scale invariant spectrum of fluctuations with a 
significant non-gaussian contribution \cite{cyclic_ng}. The latter serves as a distinct prediction to the inflationary scenarios, and is
available for test in the near future. The branes 
collide at $\phi = -\infty$ which follows their separation and running of moduli field  towards the positive value of potential, leading to a cyclic phenomena.  Thus, the  
4-dimensional effective dynamics description of the Ekpyrotic/Cyclic model is very rich and interesting.

 A primary issue in the Ekpyrotic/Cyclic model is to understand the transition from the 
contracting to the expanding branch. This transition must be non-singular and is vital to 
propagate the perturbations generated prior to the collision of the branes to the regime following the collision. 
In the 5 dimensional picture, the bulk spacetime near the collision of the branes appears as a 
compactified Milne space and it has been argued that the resulting singularity is milder than 
the conventional big bang/crunch singularity \cite{cyclic_milne}. Further, in this picture it may be possible to match the perturbations across the transition \cite{cyclic_pert1}. This however results in the mixing of the modes with a strong sensitivity of predictions to the process of transition which ignores quantum gravitational effects. Other ideas to understand this transition have been proposed including 
perturbative string theoretic treatment \cite{cyclic_alpha} and AdS/CFT correspondence \cite{tch:adscft1,cht:adscft2}. Though promising 
directions to explore, these reveal little about the transition from the contracting to the expanding branch in the 4-dimensional effective theory. If one assumes that the 4-dimensional effective theory has no inputs from a theory beyond GR, this difficulty is 
not surprising. The reason is that in 
GR such a non-singular turn around is forbidden unless matter violates null energy condition. Though it is not possible to have this violation in the Ekpyrotic/Cyclic models, 
a variant of these has been proposed in form of 
New Ekpyrotic model \cite{new_ekpy,new_ekpy1} which includes a ghost condensate. The latter is responsible for violation of null energy condition and can result in a bounce. Whether such a non-singular bounce is a generic feature of this model and is not affected by the instabilities is not an open issue. Further,  it is not clear whether the 4-dimensional effective picture with a ghost condensate as 
conjectured in this model is valid at high energies of interest \cite{new_ekpy_ghost}.

 Given that non-perturbative quantum geometric modifications and their role in singularity resolution is  well understood in the homogeneous models of loop quantum cosmology,  we now explore the effective 4-dimensional loop quantum dynamics of the moduli field to understand the transition from the contracting to the expanding branch. An underlying assumption of such an analysis 
 is the treatment of potential (\ref{cyclicpot}) as an effective one for the field $\phi$ in four spacetime 
 dimensions.  This enables us to use effective equations derived in the previous section with $\phi$ as the source of matter. An investigation on these lines was carried out earlier with an additional assumption that the spacetime be isotropic \cite{svv:cyclic}. Its main conclusions were: \\ 
 (i) the big bang/big crunch singularity is generically avoided irrespective of the choice of the parameters of the potential and initial conditions for the field $\phi$.\\
 (ii) However, it is not possible in the isotropic effective loop quantum description for the field $\phi$ to turn around from $\phi = - \infty$. Thus  a cyclic evolution as envisioned in the Ekpyrotic/Cyclic model paradigm was not possible in the isotropic loop quantum cosmology.\footnote{If the inter-brane potential is modified such that it is 
 positive for large negative values of $\phi$, then the turn around of the field and a cyclic 
 description is possible \cite{svv:cyclic}. A modification to the cyclic potential as proposed in the bicyclic scenario \cite{ffkl:negative} would hence result in a desired non-singular evolution. }

It should be noted that the absence of the turn-around of $\phi$ from large negative values is not a limitation 
of the loop quantum dynamics, but is a feature shared also by the classical theory. This can be easily seen from the 
classical Friedman equation for the isotropic model. For simplicity let us assume that 
the only contribution to the matter energy density originates from the moduli field
\be\label{classical_fried}
H^2 = \f{8 \pi G}{3} \, \rho_\phi
\ee
where 
\be\label{rho_moduli}
\rho_\phi = \f{1}{2} \, \dot \phi^2 + V(\phi) ~.
\ee 
For the moduli field to turn around it is essential that $\dot \phi$ must vanish. Using 
(\ref{rho_moduli}) in (\ref{classical_fried}) we get
\be\label{dotphimod}
\dot \phi^2 = 2 \, \left(\f{3}{8 \pi G} H^2 - V(\phi)\right) ~.
\ee 
It is clear that a turn around of the field is not possible when the potential is negative. Since in the Ekpyrotic/Cyclic model the potential is negative (and asymptotes to zero) when the branes 
are supposed to collide, $\dot \phi$ can not vanish and the moduli can not turn around. 
 Presence of additional energy density components such as 
radiation is unlikely to affect this conclusion. The reason being that in the vicinity of 
brane collision $(a \rightarrow 0)$, the dominant contribution to the energy density comes from the 
kinetic energy of the moduli.\\

Interestingly, above situation changes if the spacetime has a non-vanishing anisotropic shear.
In this case, the classical generalized Friedman equation yields
\be
\dot \phi^2 = 2 \, \left(\f{3}{8 \pi G} \left(H^2 - \f{\Sigma^2}{a^6}\right) - V(\phi)\right) ~.
\ee 
The shear term can be seen as an additional component of energy density which scales the same way as $\rho_\phi$.  Thus  depending on the strength of the shear term, it is possible that $\phi$ can turn around even when $V(\phi) <0$. However, the classical theory is singular and the above equation breaks down when spacetime curvature becomes very large. Hence,  though we get some insights on the possible role of anisotropic to alleviate the above problem, it is not possible to obtain a non-singular cyclic evolution without any additional inputs from quantum gravity or going beyond the scope of the classical theory.

Given the non-trivial role the shear term can play in the dynamics of the moduli field 
in this model, it becomes important to analyze the effective loop quantum equations 
with anisotropies.  For simplicity we assume that the only contribution to the matter density originates from the field $\phi$. 
By specifying the ${\cal H}_{\mathrm{matt}}$ corresponding to the Cyclic potential \cite{st:cyclic1} we can derive the dynamics  from Hamilton's equations for 
$c_i$, $p_i$, the moduli $\phi$ and its momentum $p_\phi$ as obtained in Sec. III. These equations can be numerically 
integrated  to obtain the 
 behavior of directional scale factors, the mean scale factor, Hubble rates, energy density and shear (which is calculated from its definition (\ref{sigma1})).

 \begin{figure}[tbh!]
\includegraphics[angle=0,width=0.55\textwidth]{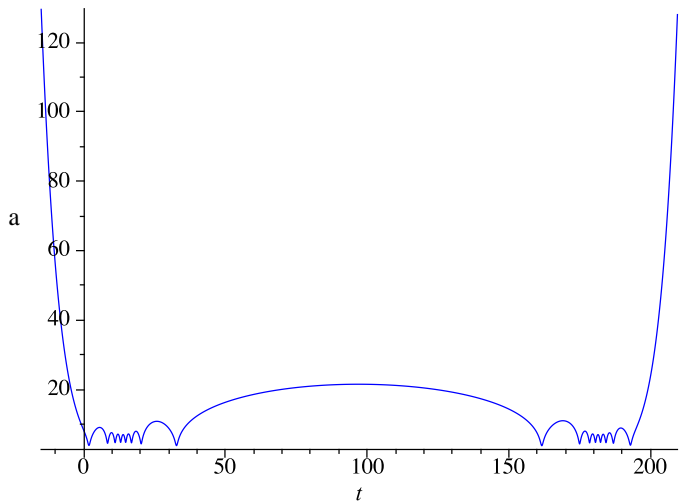}
\caption{Plot of the mean scale factor (in Planck units), $a = (a_1 a_2 a_3)^{1/3}$ is shown for the initial conditions (all values in Planck units) $\phi = 0.43$, $\dot \phi = -0.038$, $p_1 = 64$, $p_2 = 72$, $p_3 = 68$, $c_1 = -0.8$, $c_2 = -0.7$ and $\Sigma^2 = 5.80407$. The mean scale factor experiences multiple small bounces as a result of bounces of the individual scale factors $(a_i)$. Unlike the classical theory, 
there is a non-singular transition from the contracting to the expanding branch.}
\end{figure}
\begin{figure}
\includegraphics[angle=0,width=0.55\textwidth]{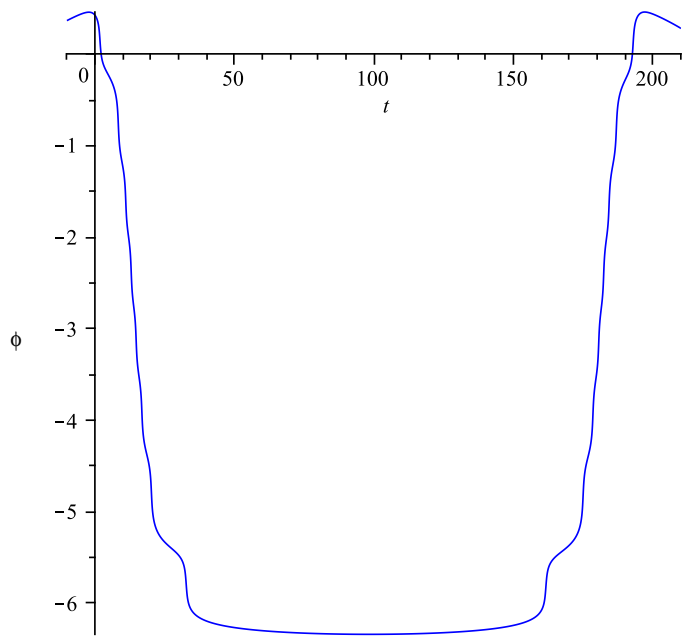}
\caption{Evolution of the moduli field in Planck units is shown for the initial conditions in Fig. 1. The moduli field starts from a positive region of the potential, rolls to the negative part and turns around in the Planck regime. After the turn around the moduli again reaches the positive part of the potential, ready for another cycle.}
\end{figure}

 \begin{figure}[tbh!]
\includegraphics[angle=0,width=1\textwidth]{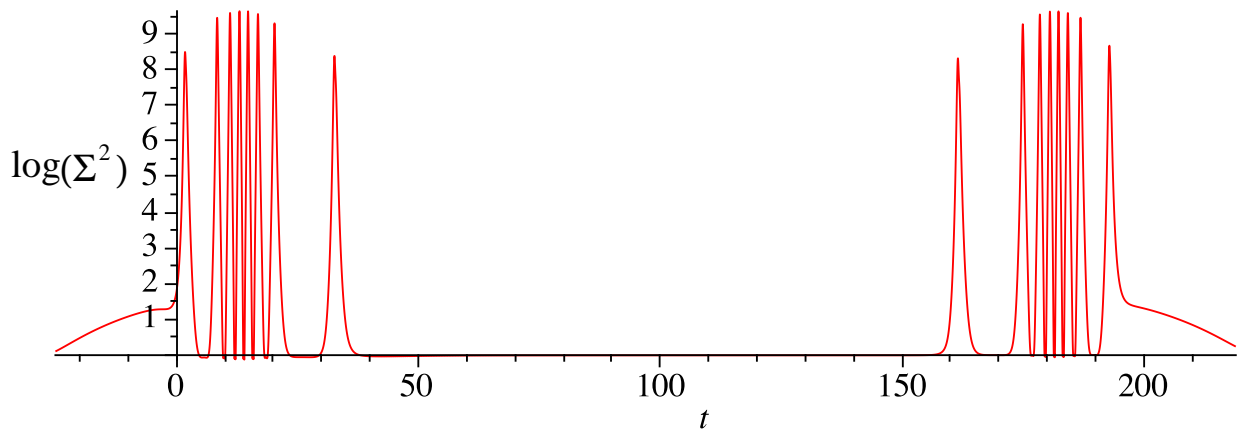}
\caption{The plot shows the variation of shear $\Sigma^2$ in the transition region from contracting to expanding branch for initial conditions in Fig. 1. Considerable variation for a short period before and after the bounce  is evident from the spikes in value of $\log(\Sigma^2)$. We also see that in the classical regimes at both small and large values of $t$, shear approaches similar values.} 
\end{figure}

\begin{figure}[tbh!]
\includegraphics[angle=0,width=0.8\textwidth]{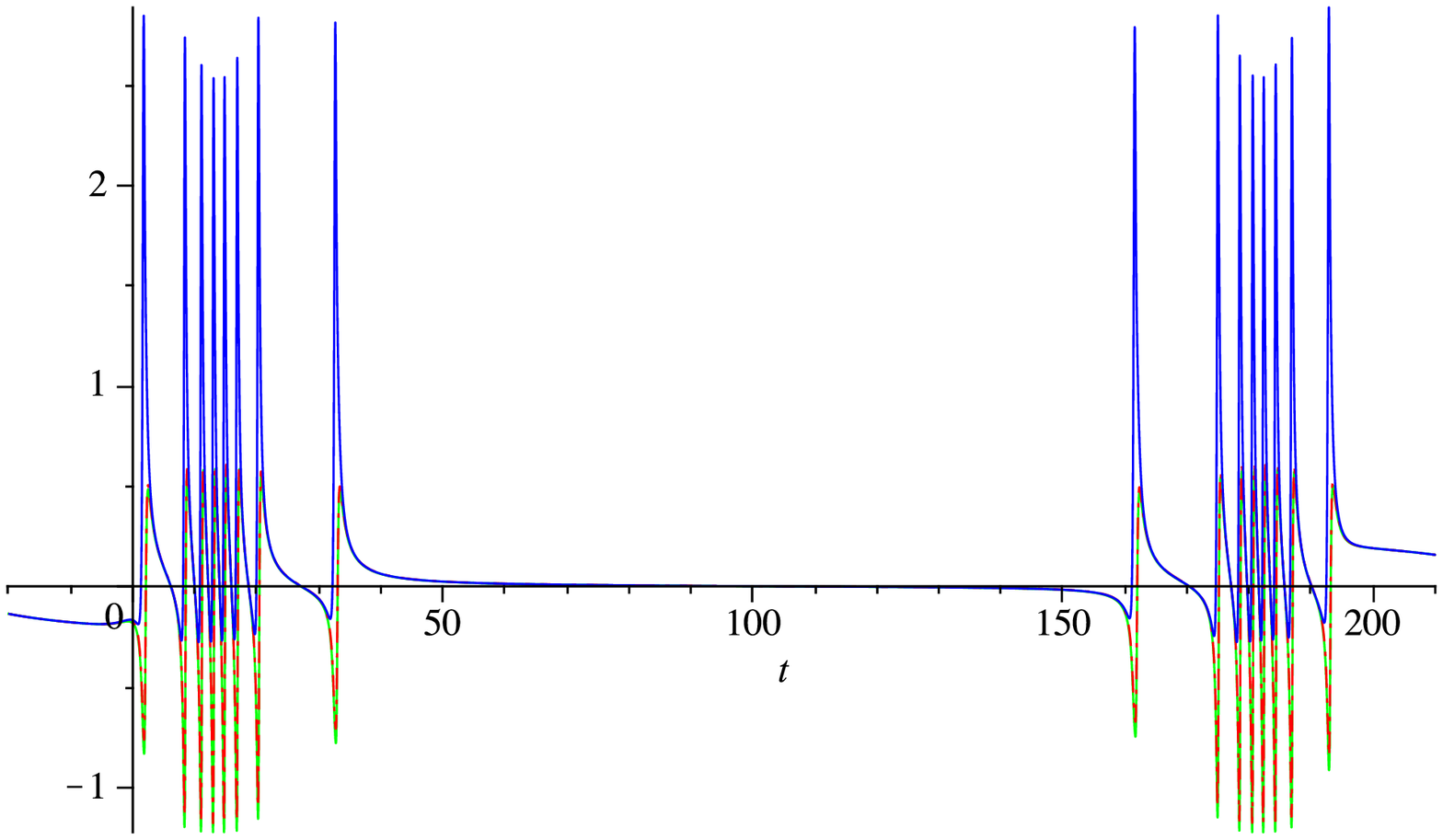}
\caption{Variation in directional Hubble rates is plotted for the initial data in Fig. 1. Green curve corresponds to $H_1$, red to $H_2$ and the  blue curve corresponds to $H_3$. 
The spikes reflect bounces and recollapses of individual scale factors in the Planck regime during the transition from the contracting to the 
expanding branch.}
\end{figure}

 We performed various numerical simulations with different initial data for the moduli field, expansion rates and  anisotropy.  The initial conditions to solve the dynamics were provided 
 for a  contracting universe with small initial anisotropy and the value of the field $\phi$ in the positive part of the Cyclic model potential.  Since the primary aim of our analysis is to investigate the resolution of singularity, we choose without the loss of generality parameters and  initial conditions independent of the consideration from the ones constrained by 
observations. 
We now illustrate some numerical results obtained from the effective dynamics.

The first set of evolution is depicted in Figs. 1-4. Parameters in the potential (\ref{cyclicpot}) were chosen as
$V_o = 0.02$, $\sigma_1 = 0.3 \sqrt{8 \pi}$ and $\sigma_2 = 0.09 \sqrt{8 \pi}$. (For simplicity we choose $G = \hbar = c = 1$). Initial conditions for the moduli field were $\phi = 0.43$ and $\dot \phi = -0.038$. Initial conditions (provided at time $t =0$) for the triad and  connection components were $p_1 = 64$, $p_2 = 72$, $p_3 = 68$, $c_1 = -0.8$, $c_2 = -0.7$ with the initial anisotropy $\Sigma^2_i = 5.80407$. (The connection component $c_3$ was determined by solving the Hamiltonian constraint ${\cal H}_{\mathrm{eff}} \approx 0$). 
These initial conditions correspond to a contraction of all the scale factors, i.e. $\dot a_1 < 0, \dot a_2 < 0$ and $\dot a_3 < 0$.
Fig. 1 shows the evolution of mean scale factor $a = (a_1 a_2 a_3)^{1/3}$. We find that the mean scale factor undergoes multiple bounces and recollapse in the Planck regime and there is a non-singular evolution from the contracting to the expanding branch. This in a sharp contrast to the classical theory where such a transition is forbidden and the contracting and expanding branches are disjoint in the 4-dimensional description. Interestingly as depicted in the Fig. 2 we also obtain a turn around of the moduli field. The moduli starts from the positive part of the potential when the universe is contracting, rolls down the negative region of potential and when the spacetime curvature reaches Planck value, it turns around and goes back to the positive part. In the subsequent evolution the classical friction term stops the moduli field and it turns around causing a contraction of the universe. Thus leading to a cyclic phenomena. In the above numerical simulation, the turn around of the moduli field 
occurs  at approximately $t \sim 100$ (in Planck units) coinciding with the midpoint of the transition from contraction to expansion of the scale factor. 

Fig. 3 shows the evolution of the shear term obtained by the numerical integration.
We find that the shear term varies significantly from its classical value in the region where the transition occurs. However, it is everywhere bounded and does not affect the occurrence of bounces. As discussed earlier, due to the property of evolution of shear 
in loop quantum cosmology, the value of shear scalar in the low curvature regime before and after the transition turns out to be the same. Also for this numerical run, the shear scalar is very small when the moduli field turns around. 
We also find that the behavior of $\Sigma^2$ consists of spikes  
in the Planck regime where the mean scale factor of the universe undergoes multiple bounces. Using the expression of shear factor (obtained from  (\ref{sigmascalar}) and (\ref{sigma1}) for loop quantum evolution), the cause of these spikes turns out to be  due to rapid variation in the Hubble rates in different directions occurring during multiple bounces. This is illustrated in Fig. 4
where in different curves  we have shown the behavior of directional Hubble rates. As can be seen the spikes in shear scalar coincide with the 
spikes in the directional Hubble rates which occur due to multiple bounces and recollapse of anisotropic scale factors in the Planck regime.\\

 \begin{figure}[tbh!]
\includegraphics[angle=0,width=0.75\textwidth]{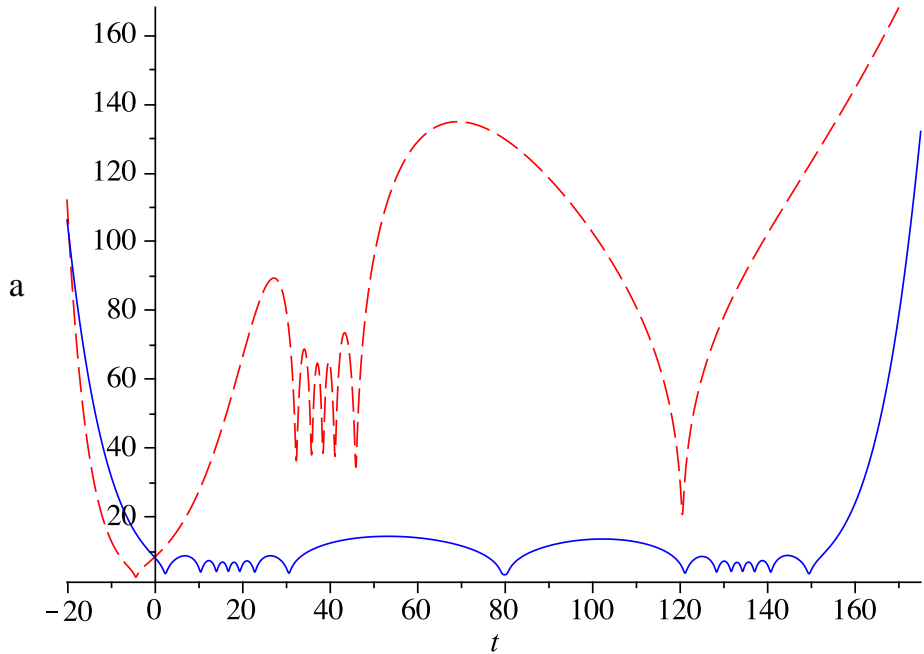}
\caption{Plot shows evolution of the scale factor in the presence (solid curve) and absence (dashed curve) of anisotropy for initial conditions 
 $\phi = 0.4$, $\dot \phi = -0.03$, $p_1 = 64$, $p_2 = 72$, $p_3 = 68$, $c_1 = -0.6$, $c_2 = -0.5$ and $\Sigma^2 = 9.2365$.
The classical singularity is avoided in both the cases. } 
\end{figure}

 \begin{figure}[tbh!]
\includegraphics[angle=0,width=0.6\textwidth]{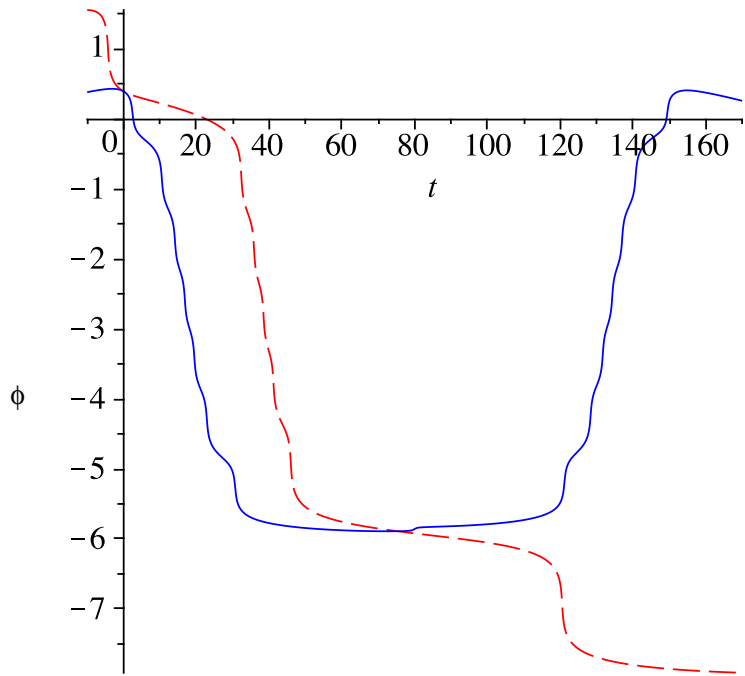}
\caption{Evolution of the moduli field is compared for initial conditions in Fig. 5 for the anisotropic (blue) and isotropic (red) model. As in Fig. 2 the moduli field turns around in the presence of anisotropies in the Planck regime and leads to a non-singular cyclic model. It fails to turn around, and approaches $\phi = - \infty$,  
in the absence of anisotropies in confirmation with results of Ref. \cite{svv:cyclic}.}
\end{figure}

\begin{figure}[tbh!]
\includegraphics[angle=0,width=0.85\textwidth]{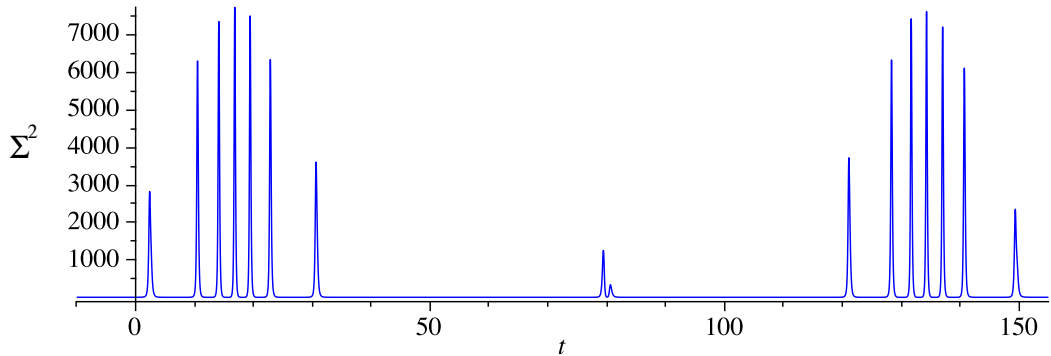}
\caption{The shear scalar $\Sigma$ is plotted for the numerical run in Fig. 5. We see that its value is preserved before and after the transition period. The shear term is bounded across the evolution with spikes corresponding to multiple bounces and recollapses of the individual scale factors.}
\end{figure}
 
 \begin{figure}[tbh!]
\includegraphics[angle=0,width=0.8\textwidth]{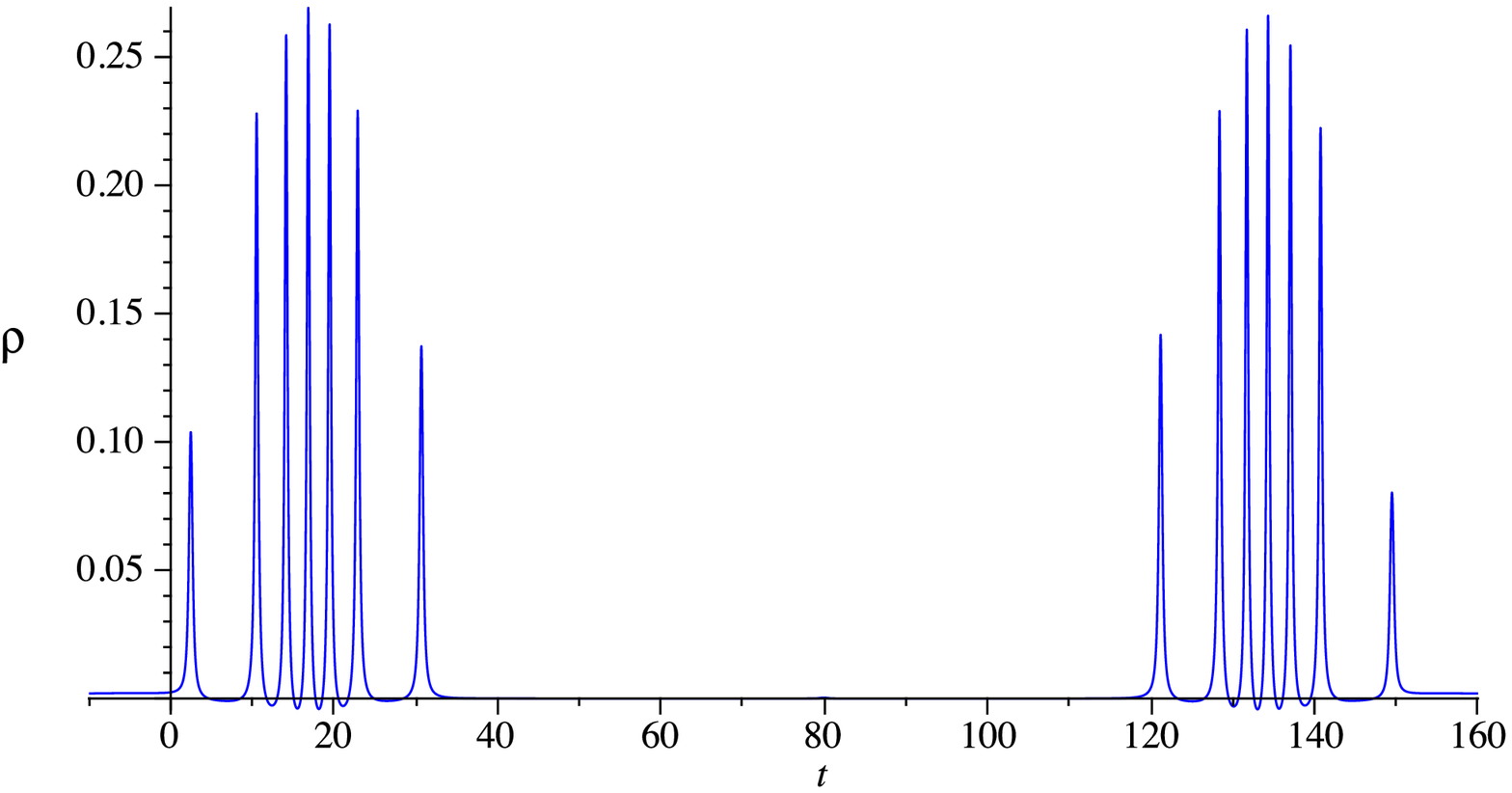}
\caption{Plot of the energy density of the moduli field in the transition regime is shown. Bounces of the individual scale factors result in spikes in its behavior. Due to presence of non-vanishing anisotropy, the energy density at these bounces does not need to saturate to its maximum allowed value.}
\end{figure}

Another example of results obtained from numerical integration are shown in Figs. 5-8.  The parameters of the potential (\ref{cyclicpot}) were chosen as $V_o = 0.01$, $\sigma_1 = 0.3 \sqrt{8 \pi}$ and $\sigma_2 = 0.09 \sqrt{8 \pi}$. The initial conditions specified at time $t=0$ were $\phi = 0.4$, $\dot \phi = -0.03$, $p_1 = 64$, $p_2 = 72$, $p_3 = 68$, $c_1 = -0.6$, $c_2 = -0.5$ and $\Sigma^2 = 9.2365$. These initial conditions correspond to 
two of the anisotropic directions contracting $(\dot a_1 < 0, \dot a_2 < 0)$ and one expanding ($\dot a_3 > 0$).
To understand the non-trivial role played by anisotropic term we also performed analysis of evolution in the isotropic loop quantum model for the same initial conditions for moduli field and 
the isotropized initial conditions for triad and connection. In Fig. 5 we have shown the evolution of the scale factor in the present model (solid curve) and the isotropic model (dashed curve). We find that the bounce of the scale factor is present both in presence and absence of 
anisotropy. The loop quantum evolution with the Cyclic potential is thus non-singular with or without anisotropy. Fig. 6 shows the evolution of 
the moduli field and we find that as for the previous case (Fig. 2) there is a turn around of the moduli field from the negative values of the 
potential to the positive part resulting in a cyclic phenomena. As before, the turn around of the field occurs at the midpoint of the transition of the mean scale factor from the contracting to expanding branch. Fig. 6 also shows that for the same initial conditions the turn around of the moduli is absent for the isotropic model. Thus, anisotropies play a very important role to obtain a non-singular viable cyclic description.

 \begin{figure}[tbh!]
\includegraphics[angle=0,width=0.4\textwidth]{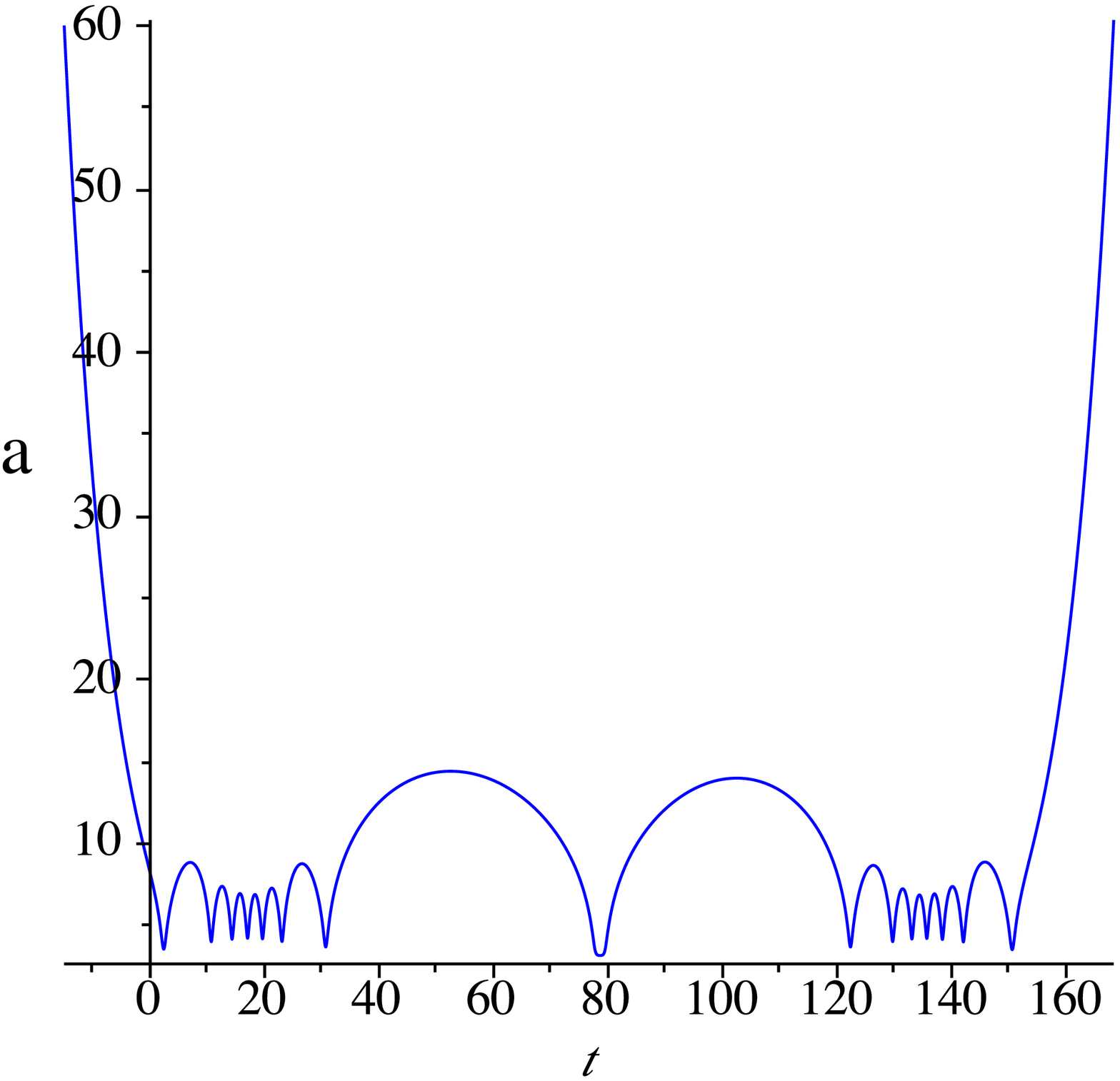}
\hskip1cm
\includegraphics[angle=0,width=0.45\textwidth]{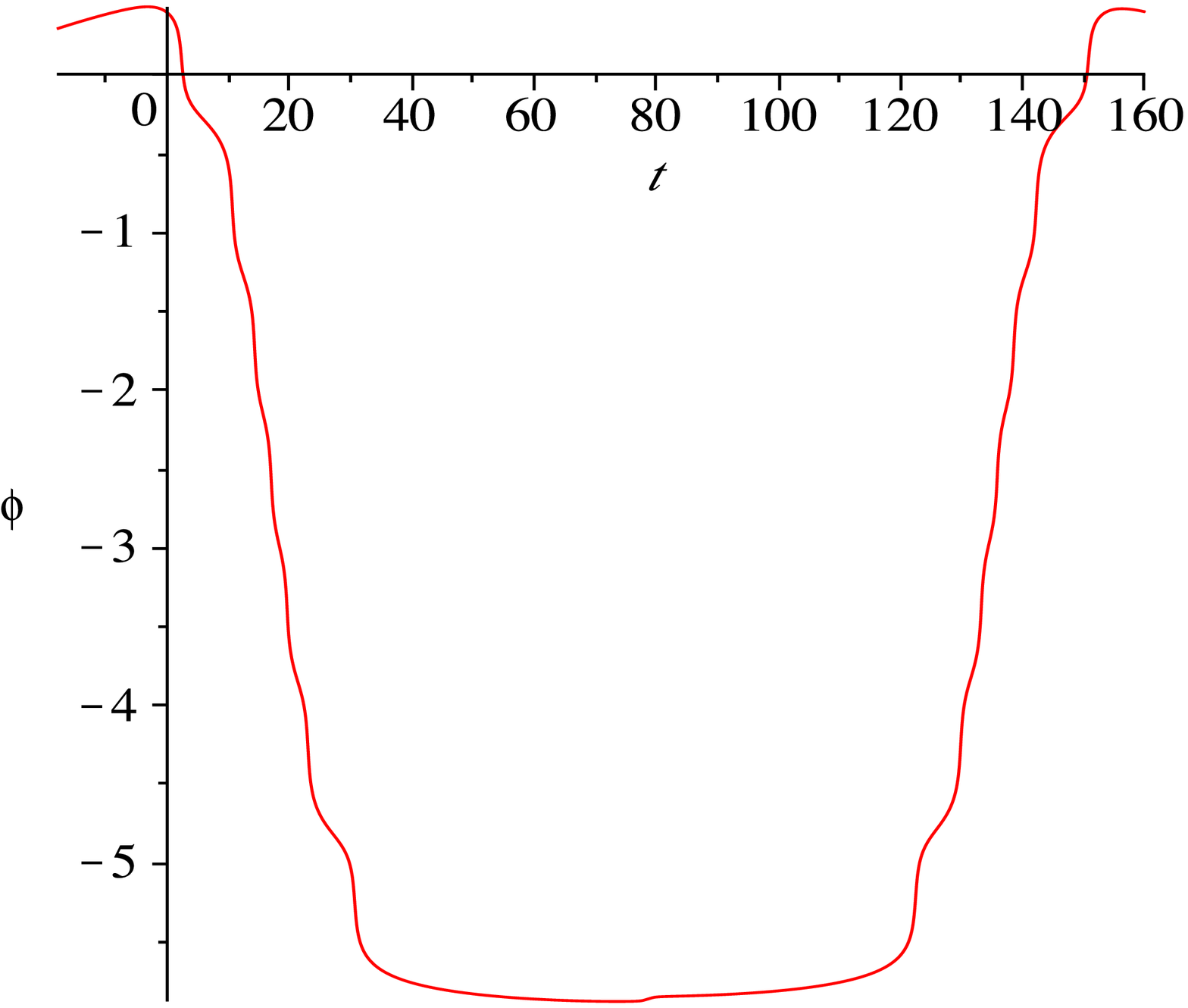}
\caption{Plots of the mean scale factor $a$ and the scalar field $\phi$ are shown for 
the initial conditions with very small anisotropies. These are $\phi = 0.4$, $\dot \phi = -0.03$, $p_1 = 64$, $p_2 = 72$, $p_3 = 68$, $c_1 = -0.6$, $c_2 = -0.532$ and $\Sigma^2 = 0.8517$.}
\end{figure}
\begin{figure}[tbh!]
\includegraphics[angle=0,width=0.4\textwidth]{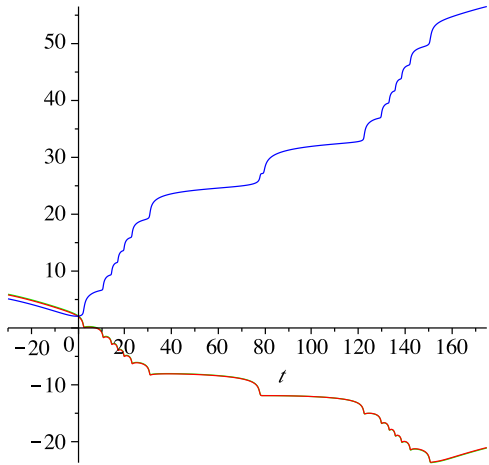}
\caption{This plot shows the evolution of different scale factors for the simulation corresponding to Fig. 9. We have plotted the logarithm of directional scale factors. The green curve (hidden behind red curve due to small anisotropy between $a_1$ and $a_2$ directions) depicts $a_1$, red curve $a_2$ and blue curve $a_3$.}
\end{figure}

\begin{figure}
\includegraphics[angle=0,width=0.5\textwidth]{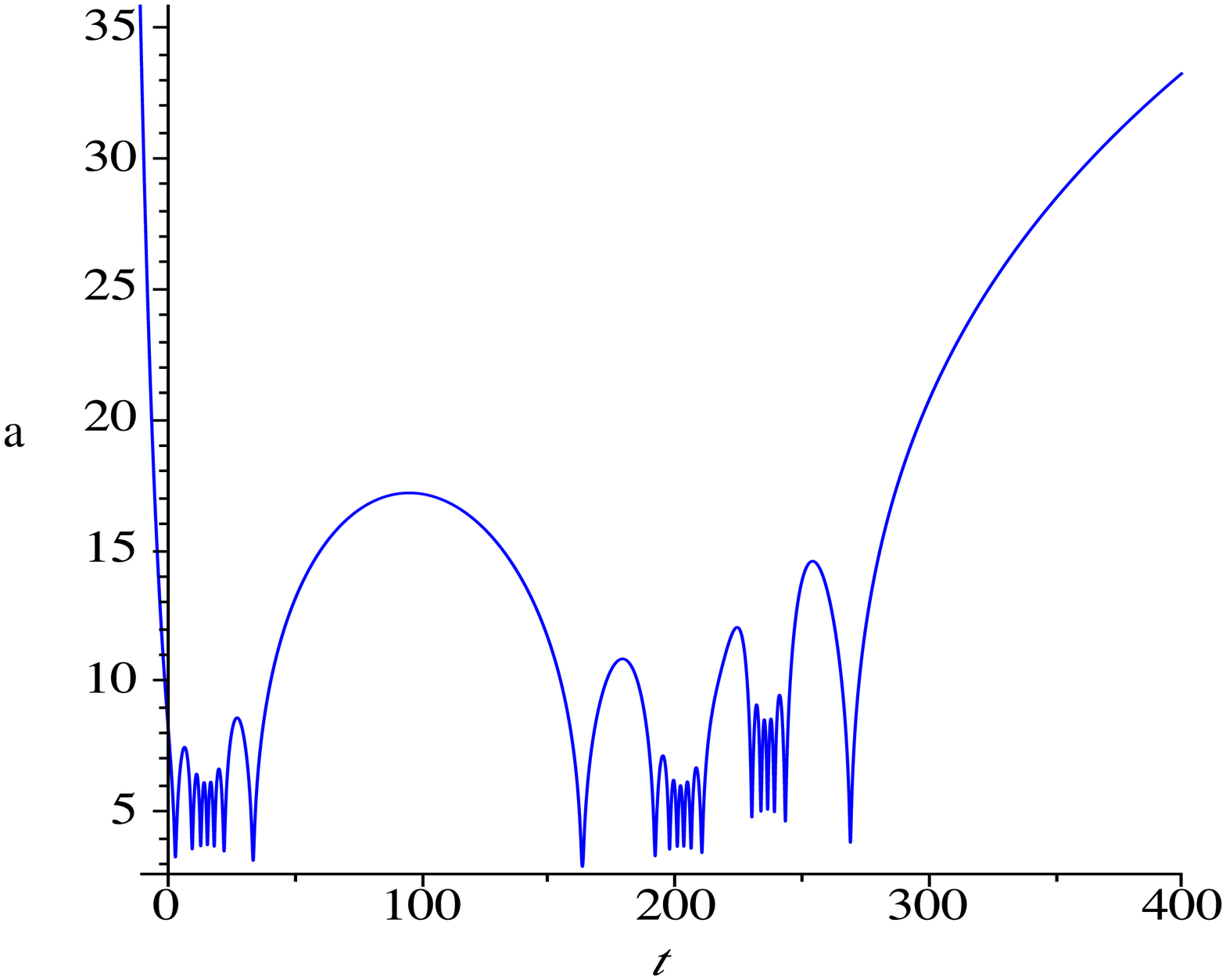}
\hskip1cm
\includegraphics[angle=0,width=0.4\textwidth]{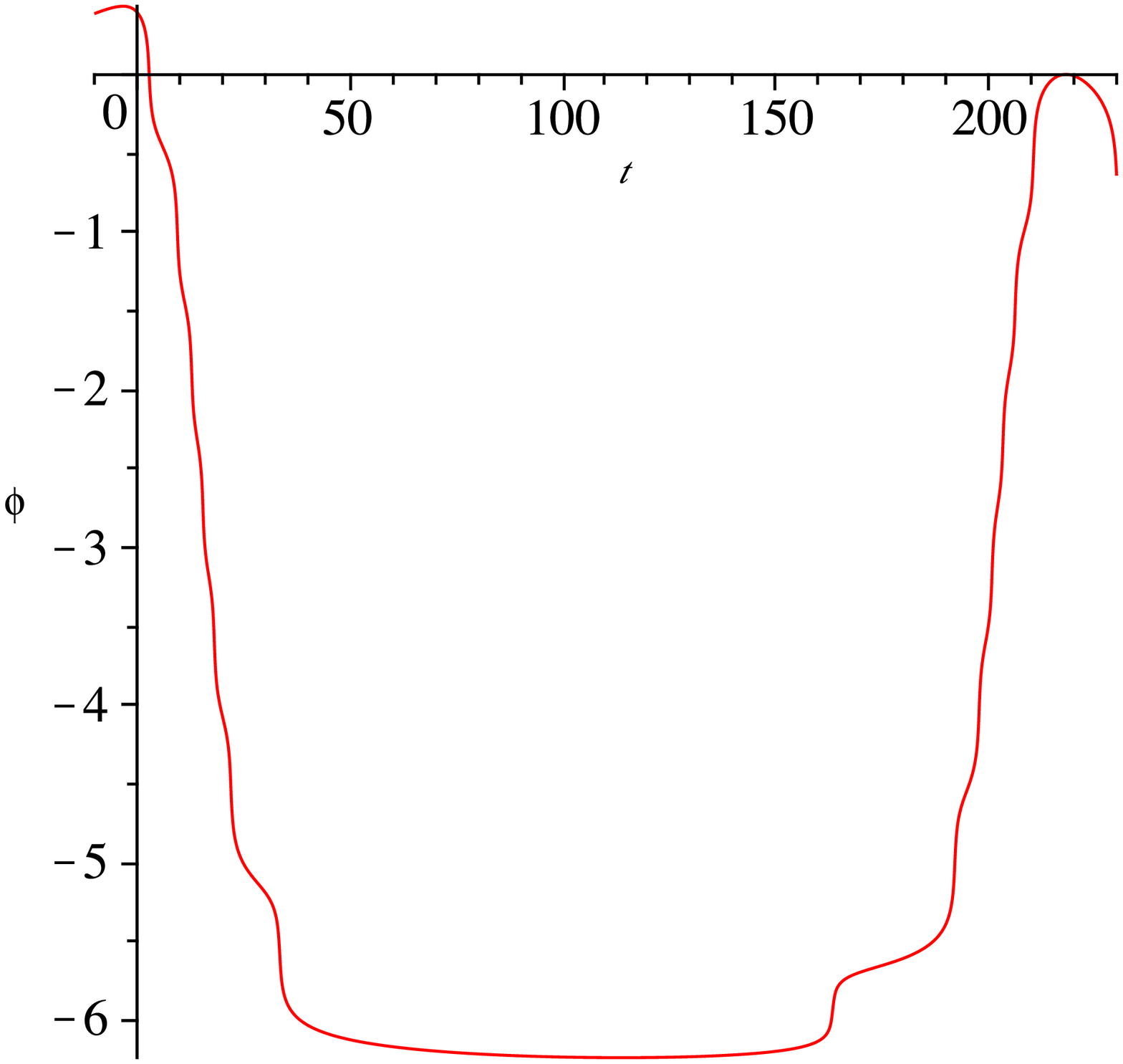}
\caption{These plots depict evolution of scale factor and scalar field for initial conditions: $\phi = 0.4$, $\dot \phi = -0.03$, $p_1 = 64$, $p_2 = 72$, $p_3 = 68$, $c_1 = -0.6$, $c_2 = -0.1$ and $\Sigma^2 = 1449.4278$. Unlike the previous cases, we find that the asymmetry in the mean scale factor and the moduli field around the bounce point is enhanced. Note that the moduli field reaches the positive value for $t \sim 220$ before starting a new cycle.}
\end{figure}

The evolution of shear and energy density are depicted in Figs. 7 and 8. We see that the shear term undergoes a significant variation in the 
region of non-singular transition but remains bounded. The spikes in the variation correspond to the spikes in the directional Hubble rates 
which occur due to multiple bounces and recollapses of the scale factors $(a_1, a_2, a_3)$. Further, its value before and after the transition is preserved. Unlike the previous case in Fig. 2, the shear scalar is not small at the turn around of the moduli field but shows a spike. 
The behavior of energy density is shown in Fig. 7 which demonstrates that at the  bounces of the directional scale factors (associated with the occurrence of spikes) the energy density does not saturate to its maximum value, a feature due to the non-vanishing anisotropy at bounces. 
 In contrast, the energy density at the bounce in the isotropic model always reaches its maximum value, $\rho_{\mathrm{crit}} = 0.41 \rho_{\mathrm{Pl}}$ at the bounce. \\
 
In both of the previous simulations, though the initial anisotropic shear is small it is 
nevertheless of the same order. An interesting question which arises is whether the non-singular transition of the scale factor and the turn around of the moduli field occurs if the shear is decreased. It is to be noted that as shown in Ref.\cite{cs:geom} 
the loop quantization of  Bianchi-I model leads to an upper bound on the 
expansion rate and anisotropic shear. These bounds are generic and independent 
of initial conditions for the matter content. The upper bound on the expansion rate 
implies that the mean scale factor would always bounce for the Cyclic model potential as the classical singularity is approached. Further, from  the discussion of the classical theory we recall that the turn around of moduli field would require appropriate choice of initial conditions of the scalar field and anisotropy. In various simulations which we performed, we found that there exists a large range of initial data for which the
turn around of moduli field can occur even for very small anisotropies.  Results from 
one such example are depicted in Figs. 9-10. We chose the same values of parameters 
as in previous simulations. The initial values are $p_1 = 64, p_2 = 72, p_3 = 68, c_1 = -0.6, c_2 = -0.532$ and $\Sigma^2 = 0.8517$. As can be seen from these plots a significant decrease 
in the initial anisotropy does not affect the bounce of the scale factor or the 
turn around of the moduli field. 

It is interesting to note that in these simulations the behavior of the mean scale factor and the scalar field seems symmetric across the middle of the transition from the contracting to the expanding branch. To understand this we first note that the weak symmetry of the variation of the scalar field in the transition regime stems from lack of interaction with any other form of matter. We expect that presence of additional matter degrees of freedom would lead to a pronounced asymmetry. Further note that in the plots we depict the mean scale factor which actually suppresses the asymmetry of the individual scale factors across the bounce.
This becomes clear if we plot the anisotropic scale factors as shown in Fig. 11 for the 
numerical simulation for Fig. 10. In Fig. 12 we show results from another numerical simulation where the asymmetry in the mean scale factor is not suppressed. The parameters are chosen as same in previous cases and the initial conditions are $p_1 = 64, p_2 = 72, p_3 = 68, c_1 = -0.6, c_2 = -0.1$ and $\Sigma^2 = 1449.4278$. As we can see the behavior of scale factor and the scalar field across the bounce is not symmetric. \\

 We summarize the main results of the numerical analysis as follows:\\
 \begin{enumerate}
\item Irrespective of the choice of initial conditions, the classical singularity at $a=0$ is generically avoided in effective dynamics of loop quantum cosmology for the Cyclic model potential.\\
 \item Starting from arbitrary anisotropic conditions, scale factors in different directions bounce when spacetime curvature becomes close to the Planck value. This causes a bounce of the mean scale factor in the Planck regime. The shear term remains bounded during the evolution and it approaches the constant classical value when spacetime curvature becomes small.\\
 \item Interestingly, it is not difficult to choose the initial conditions such that 
avoidance of the singularity is accompanied by a turn around of the moduli field in the negative regime of the potential. The moduli then rushes  towards the positive part of the potential, stops at certain positive value of the potential and roll back towards the negative value of the potential. This leads to a cyclic model of the universe. \\
\end{enumerate}

\section{Summary and Open Issues}
Ekpyrotic/Cyclic model is a very interesting paradigm for the early universe which is considered as an 
alternative to the inflationary scenarios. A key issue in this model is to understand the transition from contracting to the expanding branch
 in the 4-dimensional spacetime picture. In the 5-dimensional picture this transition corresponds to the collision between 
 boundary branes in the bulk. Though novel insights have been gained in the latter phenomenon 
 \cite{tch:adscft1,cht:adscft2,cyclic_milne,cyclic_alpha}, obtaining a non-singular transition in the 
 4-dimensional picture has remained an open issue. Understanding which  is important for various reasons including the way cosmological perturbations propagate from the contracting to the expanding branch. 
Given the Penrose-Hawking singularity theorems, a non-singular transition is not possible in the framework of classical theory. There are hopes to alleviate this problem in the classical framework by introduction of a  ghost condensate, however the approach has its own limitations \cite{new_ekpy,new_ekpy_ghost}.

Our analysis is based on the widely accepted notion that resolution of singularities involves going beyond the classical description of gravity. In particular, inputs from non-perturbative quantization of gravity may be necessary to avoid singularities. In our approach we have used the 
effective spacetime description of loop quantum cosmology to analyze the dynamics of the Ekpyrotic/Cyclic model. It is  a non-perturbative quantization of homogeneous spacetimes based on loop quantum gravity and has successfully addressed resolution of cosmological singularities in various settings \cite{aps:prl,aps:improved,acs:slqc,apsv:closed,aw:bianchi1} with a general picture of replacement of big bang with big bounce at the Planck scale. In this work we have focussed on the resolution of big bang/crunch singularity in the Ekpyrotic/Cyclic model. Such an investigation 
has been performed earlier \cite{svv:cyclic}\footnote{See also Ref. \cite{bms:cyclic} for an earlier work which ignored modifications to the 
gravitational part of the Hamiltonian.} using the assumption that the spacetime be purely isotropic. It was found that though the 
4-dimensional scale factor generically bounces in the Planck regime a viable Cyclic model is not possible due to lack of a turn around of the 
moduli field from the negative region of the Cyclic model potential.

To probe whether conclusions reached in the previous work \cite{svv:cyclic} were artifacts of ignoring anisotropic 
shear, we include them in the analysis and investigate the dynamics with the 
effective description of loop quantization of the Bianchi-I spacetime. Our analysis assumes vanishing intrinsic curvature (whose consequences are discussed below) and the moduli field as the only source of the matter energy density. The effective dynamical equations are complicated to solve analytically due to which various numerical simulations were performed with different choices of the parameter of the Cyclic model potential and initial conditions. We find that the universe undergoes a non-singular transition from the contracting to the expanding branch accompanied by 
small multiple bounces for individual scale factors in the Planck regime. In contrast to the previous results, we find that it is possible to easily choose initial conditions such that the moduli field turns around from the negative part of the potential in the Planck regime. Thus leading to a potentially viable non-singular Ekpyrotic/Cyclic model without introduction of any exotic matter. This is the novel result of our analysis.

It should be noted that the anisotropies play a very non-trivial role to obtain a non-singular cyclic model in our analysis. At first sight it seems 
perplexing because in the ekpryosis phase, which occurs when the moduli field is in the steep negative region of the potential, anisotropies become small and the universe evolves towards an isotropic phase. However as we demonstrated, even though anisotropies may become very small during the evolution, their non-zero value is important to turn around the moduli field in the negative part of the potential. Recall that 
such a turn around is not possible even for the classical theory in the absence of anisotropies. Hence the loop quantum evolution successfully leads to a turn around of the scale factor and the moduli without affecting nice features of the Ekpyrotic/Cyclic model. 

There remain several open questions which require further investigations. One of them deals with generalizing our model to include the 
intrinsic curvature of the spacetime which will require studying loop quantization of spacetimes such as Bianchi-IX. It will also help in obtaining insights on the BKL behavior in loop quantum cosmology and whether it affects the conclusion reached in this work. A second issue is to gain more analytical control on the effective dynamics of the anisotropic loop quantum cosmology. Due to its complexity it is difficult to obtain an analog of generalized Friedman equations with shear term in loop quantum cosmology, however various insights have been obtained on general features of 
the effective dynamics. These include, showing that expansion parameter and shear term are universally bounded \cite{cs:geom}. Further analytical studies in this direction on the Ekpyrotic/Cyclic model will be reported elsewhere. It will be also useful to understand the full loop quantum dynamics with the Cyclic potential and compare with the effective dynamics treatment as done presently.
Also in order to have a more realistic Ekpyrotic/Cyclic model, it is important to include radiation, study its interaction with moduli field and  its influence on the non-singular transition obtained in this work. Finally, it is an open problem to introduce 
inhomogeneities in our framework and understand the role of quantum gravitational effects on their propagation through the bounce. Recent 
results on study of Fock quantized inhomogeneous modes on the loop quantum spacetime are encouraging in this respect \cite{madrid:inhom}. It is quite possible that incorporation of these inhomogeneities may reveal subtle imprints of quantum gravity on the predictions of the Ekpyrotic/Cyclic model.

\section*{Acknowledgments}

\noindent
We are indebted to Paul Steinhardt for insightful discussions and helpful comments.
 Research at Perimeter Institute is supported
by the Government of Canada through Industry Canada and by the
Province of Ontario through the Ministry of Research \&
Innovation.





\begin{thebibliography}{10}

\bibitem{geons}
J.~A. Wheeler, ``{Geons},''
\href{http://dx.doi.org/10.1103/PhysRev.97.511}{{\em Phys. Rev.} {\bf 97}
  (1955)  511--536}.

\bibitem{pn:nature}
T.~Padmanabhan and J.~V. Narlikar, ``{Quantum conformal fluctuations in a
  singular space-time},''
\href{http://dx.doi.org/10.1038/295677a0}{{\em Nature} {\bf 295} (1982)
  677--678}.

\bibitem{ashtekar:lqc_review}
A.~Ashtekar, ``{An Introduction to Loop Quantum Gravity Through Cosmology},''
  \href{http://dx.doi.org/10.1393/ncb/i2007-10351-5}{{\em Nuovo Cim.} {\bf
  122B} (2007)  135--155},
\href{http://arxiv.org/abs/gr-qc/0702030}{{\tt arXiv:gr-qc/0702030}}.

\bibitem{aps:prl}
A.~Ashtekar, T.~Pawlowski, and P.~Singh, ``{Quantum nature of the big bang},''
  \href{http://dx.doi.org/10.1103/PhysRevLett.96.141301}{{\em Phys. Rev. Lett.}
  {\bf 96} (2006)  141301},
\href{http://arxiv.org/abs/gr-qc/0602086}{{\tt arXiv:gr-qc/0602086}}.

\bibitem{aps:mu0}
A.~Ashtekar, T.~Pawlowski, and P.~Singh, ``{Quantum nature of the big bang: An
  analytical and numerical investigation. I},''
  \href{http://dx.doi.org/10.1103/PhysRevD.73.124038}{{\em Phys. Rev.} {\bf
  D73} (2006)  124038},
\href{http://arxiv.org/abs/gr-qc/0604013}{{\tt arXiv:gr-qc/0604013}}.

\bibitem{aps:improved}
A.~Ashtekar, T.~Pawlowski, and P.~Singh, ``{Quantum nature of the big bang:
  Improved dynamics},''
  \href{http://dx.doi.org/10.1103/PhysRevD.74.084003}{{\em Phys. Rev.} {\bf
  D74} (2006)  084003},
\href{http://arxiv.org/abs/gr-qc/0607039}{{\tt arXiv:gr-qc/0607039}}.

\bibitem{kst:designingcyclic}
J.~Khoury, P.~J. Steinhardt, and N.~Turok, ``{Designing Cyclic Universe
  Models},'' \href{http://dx.doi.org/10.1103/PhysRevLett.92.031302}{{\em Phys.
  Rev. Lett.} {\bf 92} (2004)  031302},
\href{http://arxiv.org/abs/hep-th/0307132}{{\tt arXiv:hep-th/0307132}}.

\bibitem{st:cyclic1}
P.~J. Steinhardt and N.~Turok, ``{A cyclic model of the universe},''
\href{http://arxiv.org/abs/hep-th/0111030}{{\tt arXiv:hep-th/0111030}}.

\bibitem{st:cyclic2}
P.~J. Steinhardt and N.~Turok, ``{Cosmic evolution in a cyclic universe},''
  \href{http://dx.doi.org/10.1103/PhysRevD.65.126003}{{\em Phys. Rev.} {\bf
  D65} (2002)  126003},
\href{http://arxiv.org/abs/hep-th/0111098}{{\tt arXiv:hep-th/0111098}}.

\bibitem{lehners_review}
J.-L. Lehners, ``{Ekpyrotic and Cyclic Cosmology},''
  \href{http://dx.doi.org/10.1016/j.physrep.2008.06.001}{{\em Phys. Rept.} {\bf
  465} (2008)  223--263},
\href{http://arxiv.org/abs/0806.1245}{{\tt arXiv:0806.1245 [astro-ph]}}.

\bibitem{pbb_review}
M.~Gasperini and G.~Veneziano, ``{The pre-big bang scenario in string
  cosmology},'' \href{http://dx.doi.org/10.1016/S0370-1573(02)00389-7}{{\em
  Phys. Rept.} {\bf 373} (2003)  1--212},
\href{http://arxiv.org/abs/hep-th/0207130}{{\tt arXiv:hep-th/0207130}}.

\bibitem{sgas_bounce}
T.~Biswas, R.~Brandenberger, A.~Mazumdar, and W.~Siegel, ``{Non-perturbative
  gravity, Hagedorn bounce and CMB},''
  \href{http://dx.doi.org/10.1088/1475-7516/2007/12/011}{{\em JCAP} {\bf 0712}
  (2007)  011},
\href{http://arxiv.org/abs/hep-th/0610274}{{\tt arXiv:hep-th/0610274}}.

\bibitem{cyclic_pert1}
A.~J. Tolley, N.~Turok, and P.~J. Steinhardt, ``{Cosmological perturbations in
  a big crunch / big bang space-time},''
  \href{http://dx.doi.org/10.1103/PhysRevD.69.106005}{{\em Phys. Rev.} {\bf
  D69} (2004)  106005},
\href{http://arxiv.org/abs/hep-th/0306109}{{\tt arXiv:hep-th/0306109}}.

\bibitem{tch:adscft1}
N.~Turok, B.~Craps, and T.~Hertog, ``{From Big Crunch to Big Bang with
  AdS/CFT},''
\href{http://arxiv.org/abs/0711.1824}{{\tt arXiv:0711.1824 [hep-th]}}.

\bibitem{cht:adscft2}
B.~Craps, T.~Hertog, and N.~Turok, ``{Quantum Resolution of Cosmological
  Singularities using AdS/CFT},''
\href{http://arxiv.org/abs/0712.4180}{{\tt arXiv:0712.4180 [hep-th]}}.

\bibitem{ashtekar:lqc_overview}
A.~Ashtekar, ``{Loop Quantum Cosmology: An Overview},''
\href{http://arxiv.org/abs/0812.0177}{{\tt arXiv:0812.0177 [gr-qc]}}.

\bibitem{bojowald:livingrev}
M.~Bojowald, ``{Loop quantum cosmology},'' {\em Living Rev. Rel.} {\bf 8}
  (2005)  11,
\href{http://arxiv.org/abs/gr-qc/0601085}{{\tt arXiv:gr-qc/0601085}}.

\bibitem{singh:review1}
P.~Singh, ``{Transcending Big Bang in Loop Quantum Cosmology: Recent
  Advances},'' \href{http://dx.doi.org/10.1088/1742-6596/140/1/012005}{{\em J.
  Phys. Conf. Ser.} {\bf 140} (2008)  012005},
\href{http://arxiv.org/abs/0901.1301}{{\tt arXiv:0901.1301 [gr-qc]}}.

\bibitem{apsv:closed}
A.~Ashtekar, T.~Pawlowski, P.~Singh, and K.~Vandersloot, ``{Loop quantum
  cosmology of k=1 FRW models},''
  \href{http://dx.doi.org/10.1103/PhysRevD.75.024035}{{\em Phys. Rev.} {\bf
  D75} (2007)  024035},
\href{http://arxiv.org/abs/gr-qc/0612104}{{\tt arXiv:gr-qc/0612104}}.

\bibitem{skl:closed}
L.~Szulc, W.~Kaminski, and J.~Lewandowski, ``{Closed FRW model in loop quantum
  cosmology},'' \href{http://dx.doi.org/10.1088/0264-9381/24/10/008}{{\em
  Class. Quant. Grav.} {\bf 24} (2007)  2621--2636},
\href{http://arxiv.org/abs/gr-qc/0612101}{{\tt arXiv:gr-qc/0612101}}.

\bibitem{vandersloot:open}
K.~Vandersloot, ``{Loop quantum cosmology and the k = -1 RW model},''
  \href{http://dx.doi.org/10.1103/PhysRevD.75.023523}{{\em Phys. Rev.} {\bf
  D75} (2007)  023523},
\href{http://arxiv.org/abs/gr-qc/0612070}{{\tt arXiv:gr-qc/0612070}}.

\bibitem{szulc:open}
L.~Szulc, ``{Open FRW model in Loop Quantum Cosmology},''
  \href{http://dx.doi.org/10.1088/0264-9381/24/24/003}{{\em Class. Quant.
  Grav.} {\bf 24} (2007)  6191--6200},
\href{http://arxiv.org/abs/0707.1816}{{\tt arXiv:0707.1816 [gr-qc]}}.

\bibitem{acs:slqc}
A.~Ashtekar, A.~Corichi, and P.~Singh, ``{On the robustness of key features of
  loop quantum cosmology},''
  \href{http://dx.doi.org/10.1103/PhysRevD.77.024046}{{\em Phys. Rev.} {\bf
  D77} (2008)  024046},
\href{http://arxiv.org/abs/0710.3565}{{\tt arXiv:0710.3565 [gr-qc]}}.

\bibitem{aw:bianchi1}
A.~Ashtekar and E.~Wilson-Ewing, ``{Loop quantum cosmology of Bianchi I
  models},''
\href{http://arxiv.org/abs/0903.3397}{{\tt arXiv:0903.3397 [gr-qc]}}.

\bibitem{cs:recall}
A.~Corichi and P.~Singh, ``{Quantum bounce and cosmic recall},''
  \href{http://dx.doi.org/10.1103/PhysRevLett.100.161302}{{\em Phys. Rev.
  Lett.} {\bf 100} (2008)  161302},
\href{http://arxiv.org/abs/0710.4543}{{\tt arXiv:0710.4543 [gr-qc]}}.

\bibitem{ap:positivecc}
A.~Ashtekar and T.~Pawlowski, ``{Loop quantum cosmology and the positive
  cosmological constant},'' {\em In preparation} (2009)  .

\bibitem{bp:negativecc}
E.~Bentivegna and T.~Pawlowski, ``{Anti-deSitter universe dynamics in LQC},''
  \href{http://dx.doi.org/10.1103/PhysRevD.77.124025}{{\em Phys. Rev.} {\bf
  D77} (2008)  124025},
\href{http://arxiv.org/abs/0803.4446}{{\tt arXiv:0803.4446 [gr-qc]}}.

\bibitem{aps:inflaton}
A.~Ashtekar, T.~Pawlowski, and P.~Singh, ``{Pre-inflationary epoch in loop
  quantum cosmology},'' {\em In preparation} (2009)  .

\bibitem{cs:unique}
A.~Corichi and P.~Singh, ``{Is loop quantization in cosmology unique?},''
  \href{http://dx.doi.org/10.1103/PhysRevD.78.024034}{{\em Phys. Rev.} {\bf
  D78} (2008)  024034},
\href{http://arxiv.org/abs/0805.0136}{{\tt arXiv:0805.0136 [gr-qc]}}.

\bibitem{cs:geom}
A.~Corichi and P.~Singh, ``{A geometric perspective on singularity resolution
  and uniqueness in loop quantum cosmology},''
\href{http://arxiv.org/abs/0905.4949}{{\tt arXiv:0905.4949 [gr-qc]}}.

\bibitem{taveras:fried}
V.~Taveras, ``{Corrections to the Friedmann Equations from LQG for a Universe
  with a Free Scalar Field},''
  \href{http://dx.doi.org/10.1103/PhysRevD.78.064072}{{\em Phys. Rev.} {\bf
  D78} (2008)  064072},
\href{http://arxiv.org/abs/0807.3325}{{\tt arXiv:0807.3325 [gr-qc]}}.

\bibitem{willis:thesis}
J.~Willis, ``Ph. d dissertation. the pennsylvania state university,''  (2004)
  .

\bibitem{singh:effective}
P.~Singh, ``Effective equations for matter in loop quantum cosmology,'' {\em To
  appear} (2009) .

\bibitem{singh:nonsingular}
P.~Singh, ``{Are loop quantum cosmos never singular?},'' {\em Class. Quant.
  Grav} {\bf 26} (2009)  125005,
\href{http://arxiv.org/abs/0901.2750}{{\tt arXiv:0901.2750 [gr-qc]}}.

\bibitem{svv:cyclic}
P.~Singh, K.~Vandersloot, and G.~V. Vereshchagin, ``{Non-singular bouncing
  universes in loop quantum cosmology},''
  \href{http://dx.doi.org/10.1103/PhysRevD.74.043510}{{\em Phys. Rev.} {\bf
  D74} (2006)  043510},
\href{http://arxiv.org/abs/gr-qc/0606032}{{\tt arXiv:gr-qc/0606032}}.

\bibitem{ffkl:negative}
G.~N. Felder, A.~V. Frolov, L.~Kofman, and A.~V. Linde, ``{Cosmology with
  negative potentials},''
  \href{http://dx.doi.org/10.1103/PhysRevD.66.023507}{{\em Phys. Rev.} {\bf
  D66} (2002)  023507},
\href{http://arxiv.org/abs/hep-th/0202017}{{\tt arXiv:hep-th/0202017}}.

\bibitem{Lehners}
  J.~L.~Lehners and K.~S.~Stelle,
  Nucl.\ Phys.\  B {\bf 661}, 273 (2003), arXiv:hep-th/0210228.





\bibitem{cv:bianchi1}
D.-W. Chiou and K.~Vandersloot, ``{The behavior of non-linear anisotropies in
  bouncing Bianchi I models of loop quantum cosmology},''
  \href{http://dx.doi.org/10.1103/PhysRevD.76.084015}{{\em Phys. Rev.} {\bf
  D76} (2007)  084015},
\href{http://arxiv.org/abs/0707.2548}{{\tt arXiv:0707.2548 [gr-qc]}}.

\bibitem{glps:smooth}
D.~Garfinkle, W.~C. Lim, F.~Pretorius, and P.~J. Steinhardt, ``{Evolution to a
  smooth universe in an ekpyrotic contracting phase with $w > 1$},''
  \href{http://dx.doi.org/10.1103/PhysRevD.78.083537}{{\em Phys. Rev.} {\bf
  D78} (2008)  083537},
\href{http://arxiv.org/abs/0808.0542}{{\tt arXiv:0808.0542 [hep-th]}}.

\bibitem{aw:entropy}
A.~Ashtekar and E.~Wilson-Ewing, ``{The covariant entropy bound and loop
  quantum cosmology},''
\href{http://arxiv.org/abs/0805.3511}{{\tt arXiv:0805.3511 [gr-qc]}}.

\bibitem{chiou:bianchi1}
D.-W. Chiou, ``{Loop Quantum Cosmology in Bianchi Type I Models: Analytical
  Investigation},'' \href{http://dx.doi.org/10.1103/PhysRevD.75.024029}{{\em
  Phys. Rev.} {\bf D75} (2007)  024029},
\href{http://arxiv.org/abs/gr-qc/0609029}{{\tt arXiv:gr-qc/0609029}}.

\bibitem{lattice:bianchi1}
M.~Bojowald, D.~Cartin, and G.~Khanna, ``{Lattice refining loop quantum
  cosmology, anisotropic models and stability},''
  \href{http://dx.doi.org/10.1103/PhysRevD.76.064018}{{\em Phys. Rev.} {\bf
  D76} (2007)  064018},
\href{http://arxiv.org/abs/0704.1137}{{\tt arXiv:0704.1137 [gr-qc]}}.

\bibitem{cyclic_ng}
J.-L. Lehners and P.~J. Steinhardt, ``{Intuitive understanding of
  non-gaussianity in ekpyrotic and cyclic models},''
  \href{http://dx.doi.org/10.1103/PhysRevD.78.023506}{{\em Phys. Rev.} {\bf
  D78} (2008)  023506},
\href{http://arxiv.org/abs/0804.1293}{{\tt arXiv:0804.1293 [hep-th]}}.

\bibitem{cyclic_milne}
J.~Khoury, B.~A. Ovrut, N.~Seiberg, P.~J. Steinhardt, and N.~Turok, ``{From big
  crunch to big bang},''
  \href{http://dx.doi.org/10.1103/PhysRevD.65.086007}{{\em Phys. Rev.} {\bf
  D65} (2002)  086007},
\href{http://arxiv.org/abs/hep-th/0108187}{{\tt arXiv:hep-th/0108187}}.

\bibitem{cyclic_alpha}
N.~Turok, M.~Perry, and P.~J. Steinhardt, ``{M theory model of a big crunch /
  big bang transition},''
  \href{http://dx.doi.org/10.1103/PhysRevD.70.106004}{{\em Phys. Rev.} {\bf
  D70} (2004)  106004},
\href{http://arxiv.org/abs/hep-th/0408083}{{\tt arXiv:hep-th/0408083}}.

\bibitem{new_ekpy}
E.~I. Buchbinder, J.~Khoury, and B.~A. Ovrut, ``{New Ekpyrotic Cosmology},''
  \href{http://dx.doi.org/10.1103/PhysRevD.76.123503}{{\em Phys. Rev.} {\bf
  D76} (2007)  123503},
\href{http://arxiv.org/abs/hep-th/0702154}{{\tt arXiv:hep-th/0702154}}.

\bibitem{new_ekpy1}
  P.~Creminelli and L.~Senatore,
  JCAP {\bf 0711}, 010 (2007)
  [arXiv:hep-th/0702165].



\bibitem{new_ekpy_ghost}
R.~Kallosh, J.~U. Kang, A.~Linde, and V.~Mukhanov, ``{The New Ekpyrotic
  Ghost},'' \href{http://dx.doi.org/10.1088/1475-7516/2008/04/018}{{\em JCAP}
  {\bf 0804} (2008)  018},
\href{http://arxiv.org/abs/0712.2040}{{\tt arXiv:0712.2040 [hep-th]}}.

\bibitem{bms:cyclic}
M.~Bojowald, R.~Maartens, and P.~Singh, ``{Loop quantum gravity and the cyclic
  universe},'' \href{http://dx.doi.org/10.1103/PhysRevD.70.083517}{{\em Phys.
  Rev.} {\bf D70} (2004)  083517},
\href{http://arxiv.org/abs/hep-th/0407115}{{\tt arXiv:hep-th/0407115}}.

\bibitem{madrid:inhom}
D.~Brizuela, G.~A.~D. Mena~Marugan, and T.~Pawlowski, ``{Big Bounce and
  inhomogeneities},''
\href{http://arxiv.org/abs/0902.0697}{{\tt arXiv:0902.0697 [gr-qc]}}.

\end{thebibliography}
\end{document}